\documentclass[apj]{emulateapj}
\usepackage{apjfonts}

\begin{document}

\title{Dynamical Evidence for Environmental Evolution of Intermediate
  Redshift Spiral Galaxies}

\author{Sean M. Moran\altaffilmark{1}, Neil Miller\altaffilmark{1},
  Tommaso Treu\altaffilmark{2}, Richard S. Ellis\altaffilmark{1}, \&
  Graham P. Smith\altaffilmark{3}}
\altaffiltext{1}{California Institute of Technology, Department of
  Astronomy, Mail Code 105-24, Pasadena, CA
  91125, neil@caltech.edu, smm@astro.caltech.edu, rse@astro.caltech.edu}
  \altaffiltext{2}{Department of Physics, University of California, Santa Barbara,
  CA 93106, tt@physics.ucsb.edu}
  \altaffiltext{3}{School of Physics \& Astronomy, University of Birmingham,
  Edgbaston, Birmingham, B15 2TT, UK.} 

\begin{abstract}
Combining resolved optical spectroscopy with panoramic HST imaging, we
study the dynamical properties of spiral galaxies as a function of position
across two intermediate redshift clusters, and we compare the cluster
population to field galaxies in the same redshift range. By modelling
the observed rotation curves, we derive maximal rotation velocities
for 40 cluster spirals and 37 field spirals, yielding one of the
largest matched samples of cluster and field spirals at intermediate
redshift.  We construct the Tully-Fisher relation in both $V$ and
$K_S$ bands, and find that the cluster Tully-Fisher relation exhibits
significantly higher scatter than the field relation, in both $V$ and
$K_S$ bands. Under the assumption that this increased scatter is due to
an interaction with the cluster environment, we examine several
dynamical quantities (dynamical mass, mass-to-light ratio, and central
mass density) as a function of cluster environment. We find that the
central mass densities of star-forming spirals exhibit a sharp break
near the cluster Virial radius, with spirals in the cluster outskirts
exhibiting significantly lower densities.  We argue that the
lower-density spirals in the cluster outskirts, combined with the high
scatter in both $K_S$- and $V$-band TF relations, demonstrate that cluster
spirals are kinematically disturbed by their environment, even as far
as $2R_{VIR}$ from the cluster center. We propose that such
disturbances may be due to a combination of galaxy merging and harassment.

\end{abstract}


\section{INTRODUCTION}
The observed tight correlation between the rotation velocities of spiral
galaxies and their total luminosities, first noted by \citet{tully77},
has proven invaluable in helping to pin down the extragalactic
distance scale in the local universe \citep[e.g.][]{tully00}. Since
then, many authors have attempted to leverage the so-called Tully
Fisher relation to study the evolution of spiral galaxies as a
function of redshift, generally interpreting any deviation from the
local Tully Fisher (TF) relation as an evolution in luminosity.

These studies have yielded mixed results, however, with conflicting
estimates of the rate of B-band evolution as a function of redshift
\citep{bamford06, bohm04, milvangjensen03, vogt96}. 
Recently, several authors have documented many of the
systematic errors that make comparisons between studies very
difficult \citep{metevier06, nakamura06}. Indeed, \citet{nakamura06} 
argue that the most
certain method of using the TF as a measure of evolution is to
construct a large matched sample of galaxies, consisting of nearly 
equal numbers of cluster and field galaxies all measured in the same way.

In a similar manner, large matched samples such as those presented
by \citet{nakamura06} and \citet{bamford05} can also be 
effectively used to measure a
different sort of spiral galaxy evolution: that caused by infall into
a galaxy cluster. By carefully selecting galaxies across a wide range
of environments, in and out of clusters, we can gain a better
understanding of the changes in star formation and kinematics that a
spiral may undergo as it falls into a cluster potential.

In this paper, we attempt to construct such a sample out of a large
survey we are performing of two massive galaxy clusters at
intermediate redshift: Cl~0024+17 (hereafter Cl~0024) at $z=0.39$ and 
MS~0451-03 (hereafter MS~0451) at
$z=0.54$, which were selected to be complementary in their global properties, 
as part of a larger project to understand the role of the cluster environment in
galaxy evolution. We make use of high quality spiral rotation curves
determined from Keck spectroscopy to measure the Tully Fisher
relation. The sample is large enough to allow a first investigation of
the scatter of the relation as a function of
cluster-centric radius. We also study a control sample of field
galaxies in a range of redshifts centered about the cluster redshifts,
in order to asses differences between field and cluster spirals. Our
sample of 40 cluster galaxies is the largest yet reported with an
associated field sample (37 galaxies), and provides a powerful 
means to examine the effect of the cluster environment on 
infalling star-forming spirals.

In order to disentangle environmental processes affecting the
dynamics of infalling spirals and those affecting their stellar
populations, we contrast the trends in the TF relation with those
observed for integrated $V-K_S$ colors and disk mass density. The latter
quantity is constructed from rotation curves and HST scale lengths,
and is predicted to be sensitive to the strength of the `harassment' 
process \citep{moore99}.

A plan of the paper follows.
In \S~2, we describe our cluster survey and sample selection, and then
detail our procedure for deriving maximal rotation velocities in \S~3.
In \S~4, we present our results on the cluster and field Tully Fisher
relation, with an examination of other dynamical quantities in \S~5.
The Tully Fisher relation in Cl~0024 has also been studied by
\citet{metevier06}, and in \S~4 we also directly compare rotation
measurements for several galaxies in common between studies.
In \S~6 we discuss these results in light of proposed physical
mechanisms acting in the cluster environment. In this paper, we adopt a cosmology with $\Omega_m=0.3$, $\Omega_{\Lambda}=0.7$, and $H_0=70 \textrm{km s}^{-1} \textrm{Mpc}^{-1}$.

\section{OBSERVATIONS}
In this study, we leverage a large imaging and spectroscopic survey of two 
massive intermediate redshift clusters: Cl~0024 at $z=0.39$
and MS~0451 at $z=0.54$, yielding high quality spectra
of both field and cluster spirals, suitable for extracting rotation
curves. In the following, we describe our photometric and
spectroscopic observations, data reduction, and sample selection.
\subsection{Imaging}
 
We make use of {\it HST} imaging of Cl~0024
and MS~0451 from the comprehensive wide-field survey described in
\citet{tt03} and Smith et al. (2006, in
preparation). In Cl~0024, HST coverage consists of a sparsely-sampled
mosaic of 39 WFPC2 images taken in the {\it F814W} filter ($\sim I$ band),
providing coverage to a projected radius $> 5$ Mpc. MS~0451
observations were taken with the ACS, also in {\it F814W}, and 
provide contiguous coverage
within a 10Mpc$\times$10Mpc box centered on the cluster. Both sets of
observations are complete to $F814W>25$.

For Cl~0024, \citet{tt03} reported reliable morphological
classifications to $I=21.1$.  The MS~0451 observations are proportionately
deeper, and so reliable morphological classification is possible to 
the same rest frame absolute magnitude ($M_V=-19.5$). 
All galaxies brighter than this limit are classified visually
following \citet{tt03}.

We also use panoramic ground-based $K_s$-band imaging of both clusters, and
J-band imaging of Cl~0024, with the WIRC camera \citep{wilson03} on the
Hale 200" Telescope.  These data comprise a $3\times3$ mosaic of pointings,
spanning a contiguous area of $26'\times26'$ centered on each
cluster. Observations were made in 2004 November for MS~0451 and 2002 
October for Cl~0024.  The details of the observations and data 
reduction are described by \citet{smith05} and Smith et al. 
(2006, in prep.) for Cl~0024 and MS~0451,
respectively.  Point sources in the final reduced mosaics have a full
width half maximum of 0.9" and 1.0" in Cl~0024 and MS~0451 respectively, and
the $3-\sigma$ point source detection thresholds are $J=21.6$ and $K_s=19.7$ 
for Cl~0024 and $K_s=20.2$ for MS~0451.

\begin{figure}[h]
  \includegraphics[width=0.95\columnwidth]{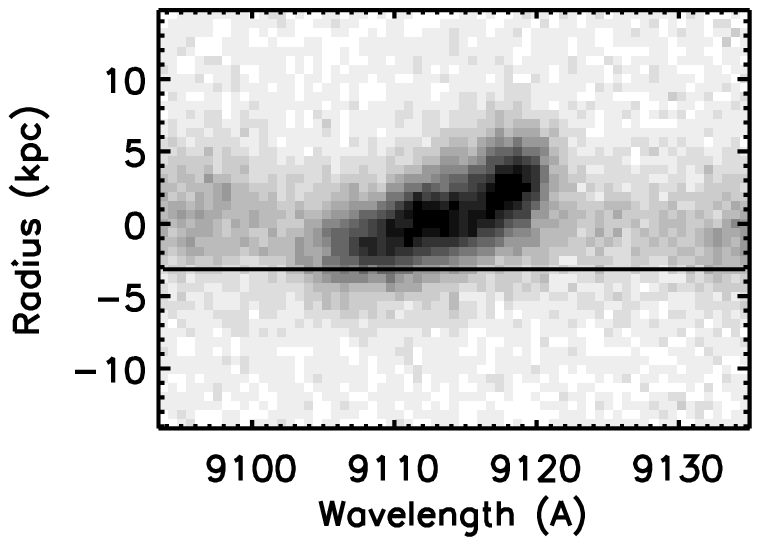}
  \includegraphics[width=0.95\columnwidth, height=2.5in]{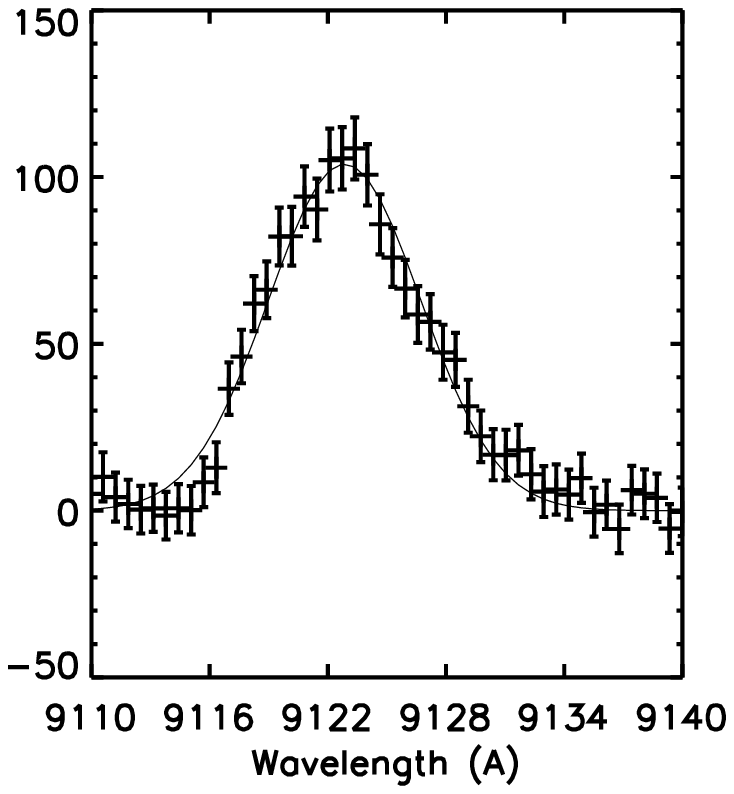}
  \caption{Top: Postage stamp image of a 2D spectrum, centered on the
    observed wavelength of H$\alpha$ for the galaxy N57426, in the field of
    MS~0451. The solid line across the image indicates the spectral
    row that is plotted in the lower panel. Bottom: One spectral
    row from the 2D spectrum. The solid curve indicates the
    Gaussian fit to this row. The stellar continuum emission 
    has been subtracted according to the method described in the text.}
  \label{spec-fig}
\end{figure}

These near-infrared data are supplemented with wide-field ground-based
optical imaging.  We make use of BVRI-band imaging of Cl~0024 with the
3.6-m Canada-France-Hawaii Telescope using the CFH12k camera \citep{cuillandre00}, full details of which are available in \citet{czoske02} and
\citet{tt03}.  MS~0451 was observed by \citet{kodama05} for
the PISCES survey through the BRI-band filters using Suprime-Cam on the
Subaru 8-m Telescope.  Full details of these data are published by \citet{kodama05}.  The CFH12k data reach $3-\sigma$ depths of $B=27.8$,
$V=26.9$, $R=26.6$ and $I=25.9$ in $\sim0\farcs9$ seeing, 
and the Suprime-Cam data reach  $3-\sigma$ depths of 
$B=28.1$, $R=27.3$, $I=25.8$ in seeing ranging from $0\farcs6$ to 
$1\arcsec$. The field of view of 
all of these optical data is well matched to the area surveyed by the 
near-infrared mosaics discussed above.

\begin{figure*}
\centering
\includegraphics*[height=7.8in]{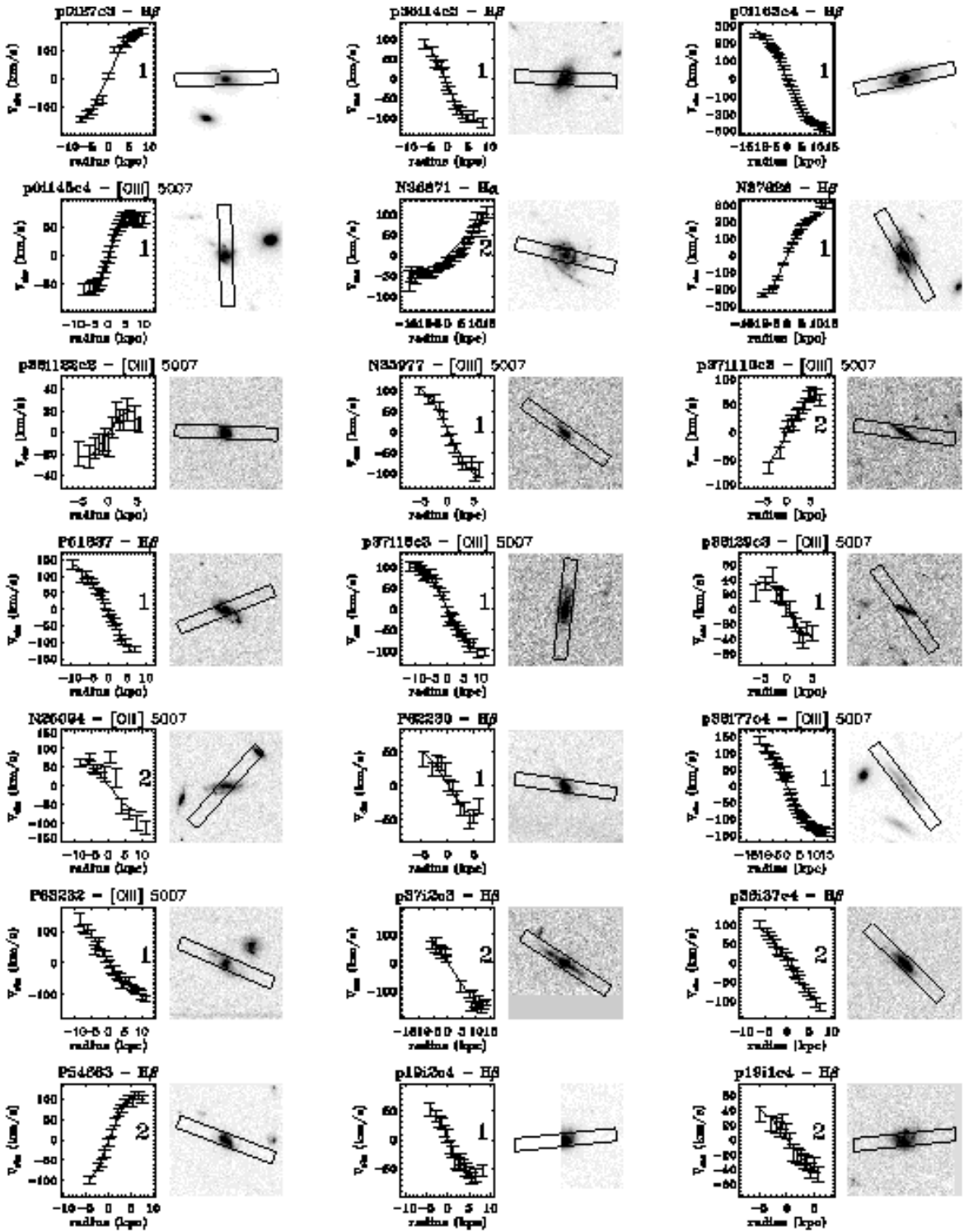}
\caption{\label{cluster_stamps} Postage stamp images and rotation
  curves for cluster spirals, arranged in order of ascending 
  projected radius, $R$, normalized by each cluster's Virial 
  radius $R_{VIR}$. {\it HST} images are $9\arcsec\time9\arcsec$, 
  in F814W.  Observed rotation curves are
  plotted as error-barred points, solid lines represent the
  best-fit rotation curve model, and the quality code
  is displayed on the plot. Radii in kpc are
  measured along the semimajor axis. In case where
  more than one emission line yielded a rotation curve, 
  we plot the best one. On the
  postage stamp images, the position of the $1\arcsec\times8\arcsec$ slit
  is indicated by the solid black box. Some slits
  are not aligned with the galaxy major axis. At
the redshift of the clusters, the spectroscopic seeing of 
$0\farcs 8$ is equivalent to $\sim 4.5 kpc$.}
\end{figure*}
\begin{figure*}
\centerline{\includegraphics*[height=7.8in]{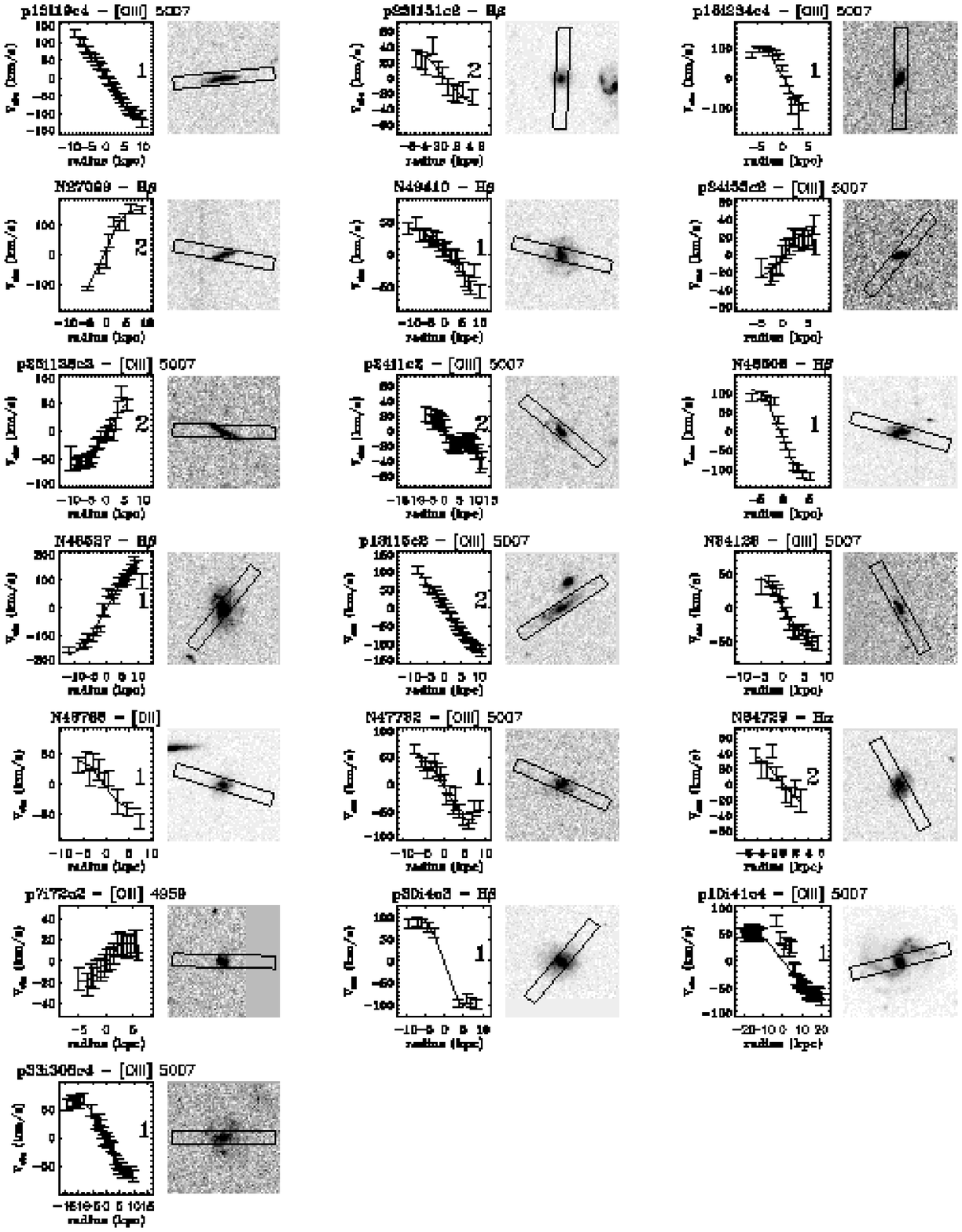}}
\centerline{Fig. 2. --- (Cont.)}
\end{figure*}

\begin{figure*}
\centering
\includegraphics*[height=7.8in]{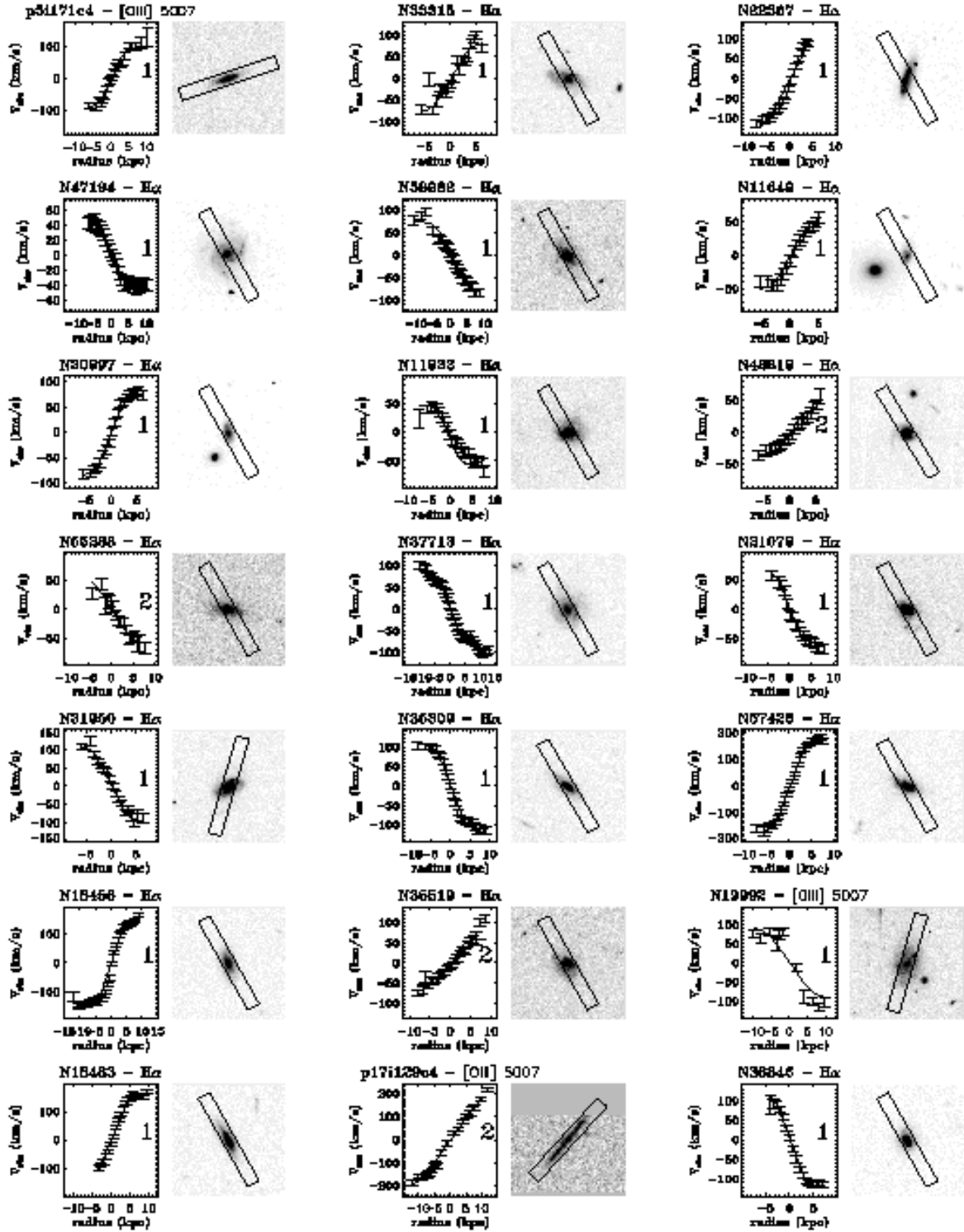}
\caption{Postage stamp images and rotation curves for all field
  spirals, arranged in ascending redshift order.\label{field_stamps}}
\end{figure*}
\begin{figure*}
\centerline{\includegraphics*[height=6.69in]{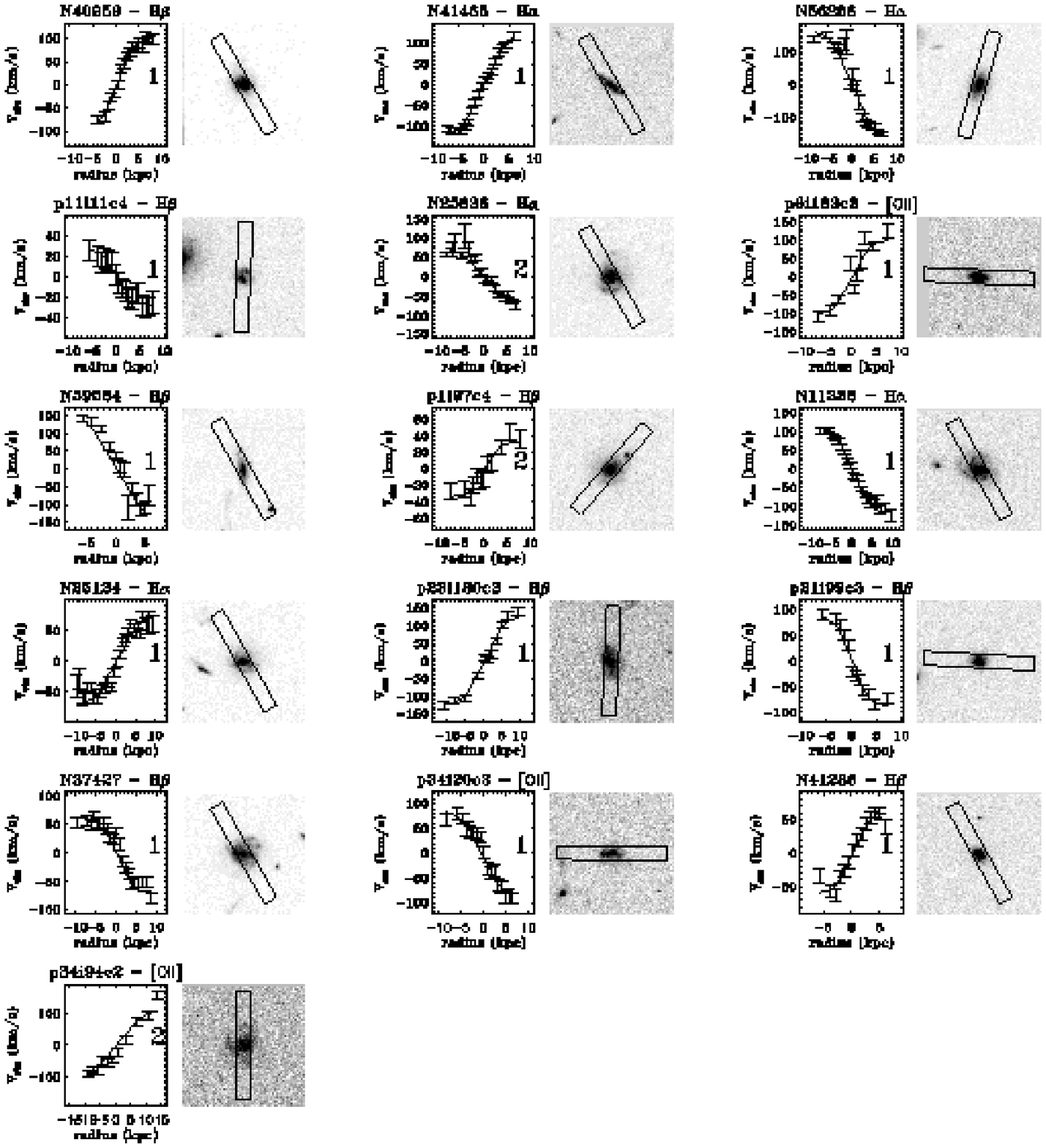}}
\centerline{Fig. 3. --- Continued.}
\end{figure*}

\subsection{Spectroscopy}
Observations with the DEIMOS spectrograph on Keck II from October 2001
to October 2005 secured spectra for over $500$ members of both
Cl~0024 ($0.373 < z < 0.402$) and MS~0451 ($0.520 < z < 0.560$). 
Details are provided in \citet{tt03},
\citet{moran05}, and Moran et al. (2007, in prep). Briefly, we observe with
$1\arcsec \times 8\arcsec$ slits, with a typical velocity resolution of 
50~km~s$^{-1}$. For each
cluster, spectral setups were chosen to span rest frame wavelengths
from $\sim3500\mbox{\AA}$ to $\sim6700\mbox{\AA}$, covering optical
emission lines [OII],[OIII], H$\beta$, and, more rarely,
H$\alpha$. Exposure times totaled 2.5hrs in Cl~0024 and 4hrs in
MS~0451.  In this spectroscopic survey, we have also obtained spectra
for over 2500 field objects, with 700 having redshifts similar to the
clusters ($0.3 < z < 0.65$).

DEIMOS data were reduced using the DEEP2 DEIMOS data reduction
pipeline \citep{davis03}, which produce sky-subtracted,
wavelength-calibrated two-dimensional spectra, suitable for identifying
and extracting extended optical emission lines. Redshifts for all
galaxies were verified by eye.

\subsection{Sample Selection}

Our current sample is drawn from the set of surveyed galaxies with
both HST imaging and available spectra.
From the HST imaging, we select candidate galaxies that
are morphologically classified as spirals (T-types 3, 4, or 5). 
For ease of comparing cluster spirals to field, we construct a matched
sample of field spirals, all with {\it HST} imaging from the Cl~0024
or MS~0451 mosaics, and selected to lie in a redshift range that
brackets the two clusters, $0.3 \le z < 0.65$.

In our spectroscopic survey, 
targets were selected randomly from an F814W-limited sample, 
to $F814W=22.5$ in the field of Cl~0024, and to $F814W=23.0$ in
MS~0451. Here, we focus on a subset of these galaxies that have 
been observed with slits aligned along the galaxy major axis, in 
order to secure resolved spectra of spirals with extended emission 
lines. The Cl~0024 campaign primarily targeted known cluster 
spirals for spectroscopy along the major axis, but serendipitous 
alignments with the major axes of field galaxies allowed us to 
include some of these in our sample. In MS~0451, we observed most galaxies 
with aligned slits, and, as a result, our 
field sample of spirals is weighted toward galaxies in the 
field around MS~0451. 

As our field sample is composed of objects in the same redshift range as the 
clusters, biases introduced by the magnitude-limited survey should affect 
both samples equally, except at the faint end of the luminosity function. 
This effect will be discussed further in \S~4.

\subsection{Source Extraction and Photometry}

Photometry was measured using SExtractor version 2.2.2 \citep{bertin96}. For
ground-based imaging, we use SExtractor in two-image mode with 
source detection performed on the ground-based
I-band images. Source detection on the HST images was performed
independently, and then matched to the ground-based 
catalog. For all imaging, we adopt magnitudes from the 
MAG\_AUTO measurement 
of SExtractor. Cl~0024 $K_S$ images are not deep enough to detect all 
cluster spirals of interest; we therefore report $K_S$ magnitudes only for 
objects detected with $>3\sigma$ significance.

The {\sl kcorrect} software v.4\_1\_2 \citep{blanton03} was used to
convert observed {\it $K_S$} and {\it F814W} fluxes to absolute
magnitudes, {\it $M_{K_S}$} (hereafter {\it $M_K$}) and {\it $M_V$}, respectively. At the
redshifts of the clusters, observed $I$-band corresponds closely to
rest-frame $V$, and so the k--corrections are small ($<0.5 mag$). 
In determining the k-corrections, we make use of all available 
ground-based photometry for each cluster field. This yields a 
better k--correction due to better
sampling of the galaxy spectral energy distributions. For field
objects at the low end of our redshift range, observed $R$-band is a 
closer match to rest-frame $V$ than observed $I$, and so for these
objects we apply a k--correction to the $R$ magnitudes to determine $M_V$.
Before determining the k--corrections, we
correct for a Galactic extinction of E({\it B--V})=0.056 or E({\it
B--V})=0.033, for Cl~0024 and MS~0451, respectively
\citep{schlegel98}. All absolute magnitudes are expressed on the AB
magnitude system.

\section{Rotation Curve Analysis}

In order to construct the Tully Fisher relations, and to study other
kinematic properties of spirals (such as mass, density, or M/L), we seek
to determine the maximum velocity of the rotation curve for each
disk. Our process involves extracting the observed rotation curve from 
the spectrum and then creating artificial rotation curves for each 
galaxy, determining the best-fit maximum velocity by $\chi^2$ fitting 
against the extracted rotation curve.  In order to do this
is important to determine various parameters about the galaxy from 
photometric data including: the position angle of the slit with
respect to the position angle of the major axis of the galaxy, the
scale length of the galaxy and the seeing when the galaxy was observed.
Along each step, we filter out the galaxies with weak spectral lines
or those which, for reasons of inclination, etc. would not be possible to fit
using our model.  We further divide the remaining fits into two subsamples, 
those with secure rotation velocities ($Q=1$), and those where the 
velocity is less certain but probably correct ($Q=2$). 
We largely follow the procedure of \citet{bohm04}, 
though several other authors have followed similar procedures
\citep[e.g.][]{metevier06, nakamura06, bamford05, vogt96}.
We have made several modifications to the procedure, detailed below.

\subsection{Extraction of Rotation Curves}

From each complete 2D spectrum, we extract postage stamps about
  the position of every emission line present, using the known redshift
  of each object to identify lines. In the top panel of 
  Figure~\ref{spec-fig}, we display an example
  postage stamp centered on H$\alpha$ for the galaxy N57426, a
  $z=0.39$ field galaxy in the vicinity of MS~0451.  As the observed
  center of H$\alpha$ emission clearly varies across the spatial
  dimension of the spectrum, the rotation in this galaxy is already apparent.
In order to determine the observed rotation curve, we fit a Gaussian function
or double Gaussian (in the case of [OII]) to each row along the spectral 
direction, as demonstrated in the lower panel of Figure~\ref{spec-fig}. 
For the double Gaussian, we
assumed a fixed separation and that the FWHM of each
Gaussian component was the same, but allowed the amplitudes to differ
independently.

We bin spectral rows together in the spatial dimension,
as necessary, to meet a signal to noise (S/N) requirement of $\sim5 (\mbox{\AA}^{-1})$
per bin. This re-binning allows us to sum up regions of the emission 
line that are too faint to fit individually, essentially trading 
spatial resolution, which is less important near the outer flat
regions of a rotation curve, for a more reliable velocity measurement. 
When rows are binned, the x position (i.e. radius) of the resulting 
velocity point is calculated by taking a S/N weighted mean of the x
positions of all the constituent spectral rows.

Each of these binned fits were checked by hand to ensure that they were 
meaningful. All extracted rotation curves for our complete sample are 
displayed in Figures~\ref{cluster_stamps} and \ref{field_stamps}. 

We found that the fit worked best when we subtracted off the continuum
to all the rows before fitting.  We measured the continuum on the 
spectrum for each spectral line separately by summing together about
50 spectral columns on either side of the emission line's center.  
By extracting and continuum subtracting over small postage stamps,
we avoid any issues that might be caused by spatial distortion in the 
spectrum.

\subsection{Surface Photometry}
From our HST imaging, we extract a $9\arcsec \times
9\arcsec$ postage stamp image of each galaxy in our sample, with the 
galaxy centered (See Figures~\ref{cluster_stamps} and
\ref{field_stamps}.). We then use the 
GALFIT software \citep{peng02} to
fit the galaxy photometry to a Sersic profile. This fit yields
estimates of the galaxy position angle on the sky, axis ratio, and 
scale length.  These parameters are necessary to determine the 
maximum velocity of the galaxy in the following step. 
While we experimented with fitting to an exponential disk plus bulge
component, we find that the scale length as derived from the 
Sersic profile fit yields the best estimate of the rotation 
curve turn-over scale length, as described below.

In local Tully-Fisher studies, internal extinction of spiral galaxies 
has been found to depend on the galaxy's inclination to
the line of sight \citep{tully85, tully98, verh01}. 
Galaxies viewed close to edge-on have a larger fraction of their 
luminosity extinguished by dust than the same galaxy would
have if viewed face on. We correct for this effect by adopting the
particularly simple form of the correction introduced by \citet{tully98}:

\begin{equation}
  A_{\lambda} = - \gamma_{\lambda} \log \left(\frac{a}{b}\right)
\end{equation}

where $a/b$ is the axis ratio of the galaxy. This formula corrects
toward the face-on case, but does not correct for additional
extinction in a face on galaxy. We choose not to apply any 
additional correction for the internal extinction of a face-on galaxy.

Following \citet{tully98}, we determine $\gamma$ in each band by
minimizing the scatter in the rest-frame color magnitude relations
$B-K_S$ vs. $M_K$ and $V-K_S$ vs. $M_K$, using the entire cluster plus field
sample together. Since the luminosity function of cluster spirals may not be
uniform across all studied environments, we ignore any luminosity 
dependence of the $\gamma$ correction, in order to avoid `fitting
away' real deviations that may be due to the cluster environment. 
We find $\gamma_B = 1.37$, $\gamma_V=1.12$, and $\gamma_K =
0.15\times\gamma_B=0.206$.  These values are consistent with
the range specified in, e.g., \citet{tully98} and \citet{verh01}.

\subsection{Model Fitting}
In order to determine the peak rotation velocity of a galaxy, we used the 
parameters obtained from GALFIT to construct an estimated velocity field 
for some maximum velocity.  We adopt a standard rotation curve 
function of the form
\begin{equation}
  V(r) = \frac{ V_{MAX} r}{(r^a + r_s^a)^{1/a}}
\end{equation}
where $r_s$ is the Sersic profile scale length as determined by
GALFIT and $a = 5$, following \citet{bohm04}. For two galaxies in our
sample, p1i97c4 and N46608, $r_S$ from GALFIT appeared to be an overestimate 
of the rotation curve
scale length, and so we manually adjusted it to achieve a better
fit. 
From the intrinsic rotation curve specified above, we 
construct a 2D velocity field by populating a grid of 
line-of-sight velocities, under the following formula:
\begin{equation}
  V_{\textrm{obs}} = V(r) \cdot cos(\phi) \cdot sin(i)
\end{equation}
where $\phi$ is the azimuth in the plane of the disk and $i$ is the 
inclination, with $i=90^\circ$ defined to be edge-on to the line of sight. 

\begin{deluxetable*}{lcccc}[t]
  \centering
  \tablewidth{0pt}
  \tablecaption{Summary of sample selection.}
  \tabletypesize{\tiny}

  \tablehead{\colhead{ } & \colhead{Cl~0024} & \colhead{MS~0451} & \colhead{Field} & \colhead{Total} \\
\colhead{ } & \colhead{$0.37<z<0.41$} & \colhead{$0.52<z<0.56$} &
\colhead{$0.3<z<0.65$} & \colhead{ }}
  \startdata
a. Spirals in the specified redshift range with {\it HST} imaging 
and DEIMOS spectra & 103 & 130 & 194 & 427 \\
b. Those with extended emission lines and aligned slits & 92 & 103 & 62 & 257 \\
c. Significant spatial extent, with a measured velocity gradient & 42 &
26 & 47 & 115 \\
d. After removing very face-on galaxies and other mis-aligned slits & 28 &
16 & 38 & 82 \\
e. Velocity uncertainty small enough & 24 & 16 & 37 & {\bf 77} \\
f. Q=1 rotation curve & 17 & 11 & 30 & 58 \\
\enddata

\end{deluxetable*}

We then convolve this velocity field with a point spread 
function (PSF) with FWHM equal to the seeing.  For our data, we used 
a fixed seeing of $0.8\arcsec$, equal to the median seeing of our 
observations. We adopt this fixed seeing correction because of the
relative insensitivity of the results to small variations in seeing; 
we find that our uncertainty
in the seeing correction affects the final $V_{MAX}$ by $<1\%$, 
which is insignificant compared to errors due to inclination or 
position angle.

Then, comparing the slit PA of our observation to the GALFIT 
estimate of the galaxy major axis angle, we place a mock 
$1\arcsec$-wide slit across
the model velocity field, at an angle reflecting the alignment
between the real slit and the galaxy major axis. Finally, at each 
position along the  length of the slit, we averaged the pixels across 
the slit width to determine an observed velocity.  
We use a $\chi^2$ minimization technique to vary the maximum 
velocity in the model to match our
observed rotation curves.  

For each observed spectrum, we estimate the position of the galaxy's spatial 
center by fitting a Gaussian function to the spatial profile of the 
2D absorption spectrum, integrated along the spectral dimension in two bands
bracketing the emission line of interest. An initial estimate 
of the velocity center is
calculated from the previously-determined redshift of each galaxy.
Both the spatial center and velocity center 
are left to be free parameters in 
the $\chi^2$ minimization. However, in no case does the best-fit spatial center 
of a $Q=1$ rotation curve differ by more than two pixels 
($\sim1.5$ kpc physical) from the calculated position, with typical offsets 
of much less than one pixel.

We first fit the rotation curve using fixed 
values for the PA and inclination, $i$.  In order to determine the 
error values on our fits, we factor in the error from the fit as well 
as computing a PA error and inclination error for the model by
running it at $\pm 10^{\circ}$ for each parameter. Especially for
galaxies that present a somewhat face-on profile, it is important to
account for this uncertainty due to PA and inclination errors, as it can
be large in some cases, and in fact causes us to discard several
emission line galaxies from our sample. We choose to vary over $\pm
10^{\circ}$ because the formal errors in the photometric fit are small
in comparison to the systematic uncertainty in measuring the PA and
inclination from the inherently asymmetric light profile of a spiral galaxy.

\subsection{Quality Control}

We began with a sample of 257 candidate spiral galaxies, each with
visible, spatially resolved emission lines. Out of this sample, we
removed 142 galaxies because the rotation curve lacked enough spatial extent
to detect a reliable turn-over, or else no significant velocity gradient 
was measured, in most cases because the galaxy is oriented nearly face-on. 
Additionally, we removed 33 galaxies because the spectroscopic slit was 
too misaligned, or because the galaxy appeared too face-on to estimate 
the direction of its major axis. 

After fitting models to the rotation curves of our candidate spirals,
and culling bad fits from our sample as described above, we remove 
five additional objects with highly uncertain 
velocities, $\Delta Log(2V_{MAX}) > 0.2$. Our final sample then consists 
of 37 field spirals and 40
cluster spirals (24 from Cl~0024). The observed rotation curves, model
fits, and images of these galaxies are presented in postage stamp form
in Figures~\ref{cluster_stamps} (cluster galaxies) and
~\ref{field_stamps} (field galaxies). 

We further divide this sample into two quality classes: $Q=1$
rotation curves have turnovers detected on both sides of the curve, 
with a model fit that accurately matches the turnover at each
end. $Q=2$ curves, about 25$\%$ of the total, only show a turnover at
one end, or show other signs of an uncertain fit to the model. 
We do not simply throw out all galaxies with signs of disturbed
kinematics, but rather keep them in the $Q=2$ sample, as 
they are of considerable interest for our study of
possible interactions with the cluster environment. However, the
requirement that we identify a reasonably secure value of $V_{MAX}$ must
necessarily exclude some number of spirals with highly disturbed
rotation curves.

We summarize the sample selection process for both clusters and the field 
in Table~1. We note that we remove roughly equal fractions of galaxies 
from the cluster and field samples, at each step of the process. The
two notable exceptions have ready explanations: In step b.), a larger 
fraction of field galaxies than cluster galaxies are removed due to
misaligned slits, because we did not consistently observe field
galaxies with aligned slits. Similarly, in step c), a low fraction of MS~0451
cluster galaxies exhibited large enough spatial extent in their
rotation curves, due to an observed suppression or lack of star formation across
this massive cluster (Moran et al. 2007, in prep).
Basic data, as well as
extinction-corrected magnitudes and velocities for all 77 objects of
our main sample are listed in Table~3.

\section{The Tully Fisher Relation}

\begin{figure*}
  \centering
  \includegraphics*[width=2\columnwidth]{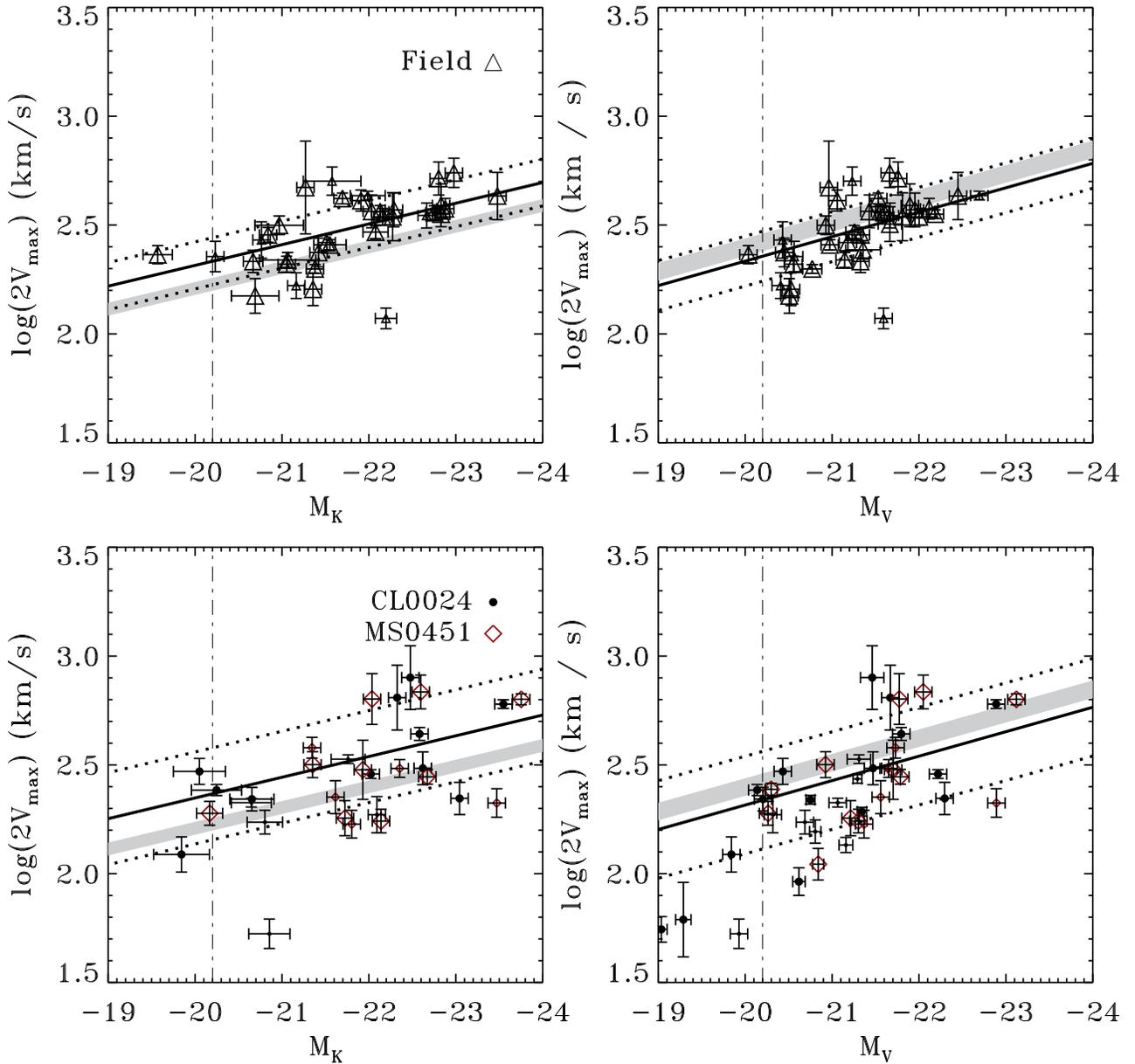}
  \caption{V--band (right) and K--band (left) Tully-Fisher
  relations for cluster and field
  spirals. The field relation is plotted in the top two panels, and 
  cluster galaxies are plotted on the bottom. Cl~0024 members are
  marked as black circles, while MS~0451 members are indicated by red
  diamonds. Shaded regions indicate the $1-\sigma$ scatter of the
  local Tully-Fisher relation, from Verheijen (2001). We adopt their R-band
  relation to compare to our V-band relation, neglecting any small correction
  to account for the different filters. 
  In each panel, the thick solid line represents the inverse--fitted TF
  relation for those points (but fixing the slope to the local value), 
  and the dotted lines represent the $1\sigma$ scatter about the
  mean. Dash-dotted line indicates where we apply a cut in magnitude
  for this calculation. Symbols are as indicated by the legend; small
  symbols indicate $Q=2$ rotation curves, and large symbols are $Q=1$.
    }
  \label{combinedTF}
\end{figure*}

In Figure~\ref{combinedTF}, we plot the Tully Fisher relation for 
both cluster and field galaxies, in both rest frame $K_S$ and $V$
bands (expressed as absolute magnitudes $M_K$ and $M_V$, respectively).
Shaded regions indicate the $\pm1-\sigma$ scatter of the local TF from
\citet{verh01}. Solid lines indicate the best fit TF zero point 
for each of our subsamples, with the slope fixed to the local 
values from \citet{verh01}, adopting their RC/FD sample, which includes
only galaxies where a turnover in the rotation curve is seen. 
Dotted lines indicate the $1-\sigma$ RMS 
scatter of the relation about the mean. Zero points are calculated by finding 
the bi-weight mean of the residuals about the local TF.  To minimize
bias, zero point and scatter are calculated only from $Q=1$ rotation 
curves. We further impose an absolute magnitude cut of $M_V\le -20.2$ and
$M_K\le -20.2$, to eliminate bias due to the differing magnitude
distributions between the cluster and field samples.

For purposes of determining the intrinsic scatter in the TF
relation for each of our samples, we also perform a least-squares fit
to find the best-fit parameters of the TF relation, weighting each
point by the measurement uncertainties in both $Log(2V_{MAX})$ and
absolute magnitude.  We follow other authors \citep[e.g.][]{bamford06,
metevier06, nakamura06} and adopt $Log(2V_{MAX})$ as the dependent
variable in the fit, such that
\begin{equation}
Log(2V_{MAX})=a+b*M_{V,K}
\end{equation} 
where we fit for intercept {\it a} and slope {\it b}.  This is the
so-called `inverse TF', which is less sensitive to bias due to
luminosity incompleteness \citep{willick94, schechter80}. For each TF
relation, cluster and field, $V$ and $K_S$ bands, the intrinsic scatter is
that portion of the measured scatter that cannot be explained by
measurement error. We estimate the intrinsic scatter by considering
the reduced $\chi^2$ statistic. Following \citet{bamford06}, we
iteratively determine the scatter that we need to add to our
measurement errors in order to achieve $\chi^2_r=1$. The best inverse
fit parameters for all four subsamples are listed in Table~2. We note
that the zero points and slopes are indistinguishable between cluster
and field; as we will discuss below, only the scatter in the relation
differs between cluster and field.

Field galaxies show a tight TF relation in both bands, with slope
consistent with the local relation \citep{verh01}.  We find an
intrinsic scatter of 0.35 mag in $V$ and 0.5 mag in $K_S$, again
restricting ourselves to $Q=1$ rotation curves brighter than our
magnitude cut. The seemingly
higher intrinsic scatter in $K_S$ is at odds with the expectation that
lower dust extinction in the $K_S$-band should make the TF tighter than
in bluer bands, and our result seems to indicate that we underestimate
the measurement uncertainties.  However, the shallower slope of the
$K_S$-band TF causes its scatter, when expressed in magnitudes, to be more
sensitive to small errors in the measured $V_{MAX}$. In fact,
expressed in terms of $\Delta Log(V_{MAX})$, the field sample scatter
in $V$ and $K_S$ are indistinguishable, yielding $0.08\pm0.01$
and $0.07\pm0.01$, respectively.
As measurement uncertainties are generally largest in the
$V_{MAX}$ direction, this simply indicates that absolute magnitude is
more properly the independent variable in our TF relation. 
In the following we will preferentially express
the TF scatter in terms of $\Delta Log(2V_{MAX})$, except when
comparing to other authors.

The 0.35 mag scatter we find in $V$ is comparable to the 
0.38 mag R-band scatter reported by \citet{verh01} for nearby
galaxies. Our $K_S$-band scatter of 0.5 mag is about 50$\%$ larger than
their reported 0.31 mag. However, \citet{kff02} have suggested that, because 
local studies tend to weed out kinematically irregular
galaxies, the true scatter, if a more representative sample of spirals
is selected, could actually be much higher. At intermediate redshift,
small irregularities in rotation curves are harder to detect, due to
limited spatial resolution, and so a higher measured scatter might
reasonably be expected.

\begin{deluxetable}{lcccc}
  \centering
  \tablewidth{0pt}
  \tablecaption{Inverse fits to Tully-Fisher relation.}
  \tabletypesize{\tiny}

  \tablehead{\colhead{Sample } & \colhead{a} & \colhead{b} &
    \colhead{} & \colhead{} \\
\colhead{} & \colhead{RMS} & \colhead{RMS} &
    \colhead{($\Delta Log(2V_{MAX})$)} & \colhead{($\Delta M_{K,V}$)}}
  \startdata
Field $K_S$ & $-0.7\pm0.3$ & $-0.14\pm0.01$ & $0.07\pm0.01$ & $0.5\pm0.05$\\
Cluster $K_S$ & $-0.5\pm0.2$ & $-0.14\pm0.01$ & $0.17\pm0.01$ & $1.18\pm0.05$\\
Field $V$ & $-2.1\pm0.4$ & $-0.22\pm0.02$ & $0.08\pm0.01$ & $0.35\pm0.05$ \\
Cluster $V$ & $-1.9\pm0.2$ & $-0.20\pm0.01$ & $0.19\pm0.01$& $0.93\pm0.05$\\
\enddata

\end{deluxetable}

Even so, the scatter we measure for the field TF is significantly
lower than has been found by other authors at intermediate redshift
\citep[e.g.][]{nakamura06, bohm04, bamford06}. In a recent large study
of 89 field galaxies, \citet{bamford06} measure an intrinsic scatter
in the B-band TF of 0.9 mag, significantly larger than our measured V-band
scatter, though they include galaxies across a larger redshift range. 
In the redder bands that we measure, we can see that a
tight TF relation still exists in the field at lookback times of over
5~Gyr.

In stark contrast to the field TF, the relation that we measure for 
cluster spirals (lower panels of Figure~\ref{combinedTF}) 
shows a remarkably high scatter, $\Delta Log(V_{MAX})=0.19\pm0.01$
(0.93 mag) and $0.17\pm0.01$ (1.18 mag) in $V$ and $K_S$, respectively.
These values are each more than twice as large as the scatter in our field
sample. This cannot be understood in terms of higher measurement 
error in the cluster sample, as the two samples were selected in the same 
way from the same parent data set, and we have restricted the analysis
to only the highest quality rotation curves. When we include all 77 rotation
curves, we observe the same difference between cluster and field, but with
overall higher measurements of scatter.

\subsection{A Comparison to Independent Measurements}
Recently, \citet{metevier06} have also published a Tully-Fisher
relation for the cluster Cl~0024, examining rotation curves of 15
spirals. Four of the galaxies in their sample are in common with our
own, allowing us for the first time to evaluate the agreement
between repeat observations and independent analysis of intermediate 
redshift spiral rotation curves. In our sample,
galaxies p0i27c3, p0i145c4, p0i163c4, and p19i2c4 correspond to 
their TFR05, TFR07, TFR10, and TFR12, respectively. 

Visual comparison of their observed rotation
curves to our own (Figure~\ref{cluster_stamps}) indicates that they
are of comparable quality, but with some differences in rotation curve
extent. Comparing our estimates
of $V_{MAX}$ to their $V_{ARC}$, we find an RMS difference of
$\Delta V/V =34\%\pm17\%$, with individual measurements differing by as much as
$100 km s^{-1}$. Three out of four measurements differ by more than
$2\sigma$. 

In \citet{metevier06}, $V_{ARC}$ is conceptually identical to
our $V_{MAX}$: both attempt to measure the broad flat part of each
spiral's rotation curve. Furthermore, our procedure for modeling the
rotation curve follows steps very similar to their GAUSS2D code. 
However, we each adopt slightly different rotation curve functions; 
they use an arctan function to approximate $V(R)$, while we use the 
function given in Equation~2, adopted from \citet{bohm04}. 

To test the effect of adopting a different rotation curve
function, we re-run our model fits for the four galaxies in common 
with \citet{metevier06}, this time fitting the observed rotation
curves to an arctan function, and adopting the best-fit scale lengths
from their paper. We find that adopting their rotation
curve function brings two of our four velocity measurements into
agreement. For the two other objects, variations in the observed
rotation curves may explain the discrepancy. In p0i163c4/TF10, 
\citet{metevier06} uncover a downturn in the rotation curve at 
high radius, which is not reached by our own data. Conversely, for 
object p0i145c4/TF12, our
rotation curve extends to larger radius and reveals that the velocity
continues to increase beyond the end of the curve measured 
by \citet{metevier06}.

\begin{figure}[t]
  \includegraphics[width=\columnwidth]{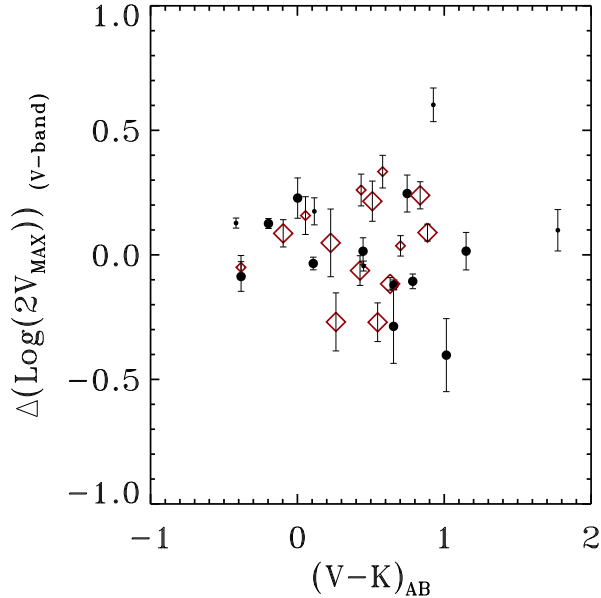}
  \caption{The residuals from the local $V$-band Tully-Fisher relation, in the
    $\log (2V_{MAX})$ direction, versus rest $(V-K_S)_{AB}$
    color. Magnitudes are corrected for inclination-dependent internal
    extinction, as discussed in the text. Symbols are coded as in Figure~\ref{combinedTF}.
  }
  \label{diffTFcolor}
\end{figure}

\begin{figure*}[t]
\centering
  \includegraphics*[width=2\columnwidth]{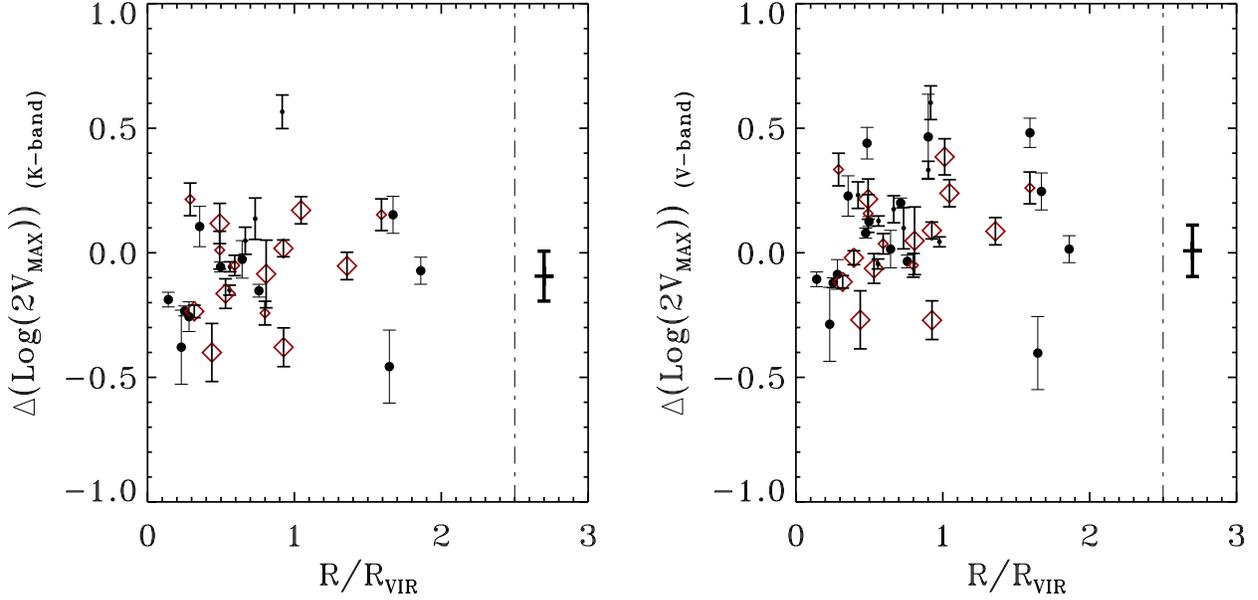}
  \caption{The residuals from the local Tully-Fisher relation
    as a function of normalized projected radius,
    $R/R_{VIR}$. Symbols are coded as in Figure~\ref{combinedTF}.
    Black crosses to the right of the
    dash-dotted line indicate the median and RMS residuals of the field TF. 
    Positive residuals indicate that
    a galaxy is overluminous for its measured velocity, or,
    alternatively, has an anomalously low $V_{MAX}$ given its
    luminosity. 
  }
  \label{diffTF}
\end{figure*}

It is striking that such a small difference in the choice of rotation
curve function can yield such a large difference in the resulting
velocity value. When comparing their results to other Tully-Fisher
studies, \citet{metevier06} take pains to comprehensively account 
for many of the systematic differences between studies, most having
to do with the way velocities were measured or defined. They show
that these small differences greatly affect estimates of the average
luminosity evolution of spirals as a function of redshift. Because of
the difficulty in comparing TF relations across samples, and the
additional, previously under-appreciated systematic arising from the
choice of rotation curve function, we do not attempt in this paper 
to make a rigorous estimate of the luminosity evolution implied 
by our TF zero point.

In fact, generally, and for our present goal of studying environmental 
influences 
on spiral galaxies, these variations between studies highlight the 
importance of matched samples of cluster and field spirals.  
Our current matched sample, which contains the largest number of
cluster galaxies so far, allows us to move beyond the Tully Fisher 
relation, to study directly the kinematics of spirals galaxies in the 
hope of uncovering the source of the large scatter seen in the 
cluster TF relation.

\section{Trends in Spiral Masses, Densities, and M/L}

Movement of a galaxy in the TF plane can be caused by various effects. Increased star formation or dustiness  moves a galaxy along
the luminosity axis, while increased total mass, changes to the
radial mass profile, or other kinematic disturbances can alter a
galaxy's measured $V_{MAX}$. In this section, we attempt to identify
the source of the high scatter observed in the cluster TF, compared to
the field in the same redshift range. Because our cluster sample spans
a large range in cluster-centric radius, it makes sense to examine the
residuals of the TF and other dynamical characteristics of the cluster
spirals as a function of environment within the clusters.

\begin{figure*}[t]
  \includegraphics*[width=0.95\columnwidth]{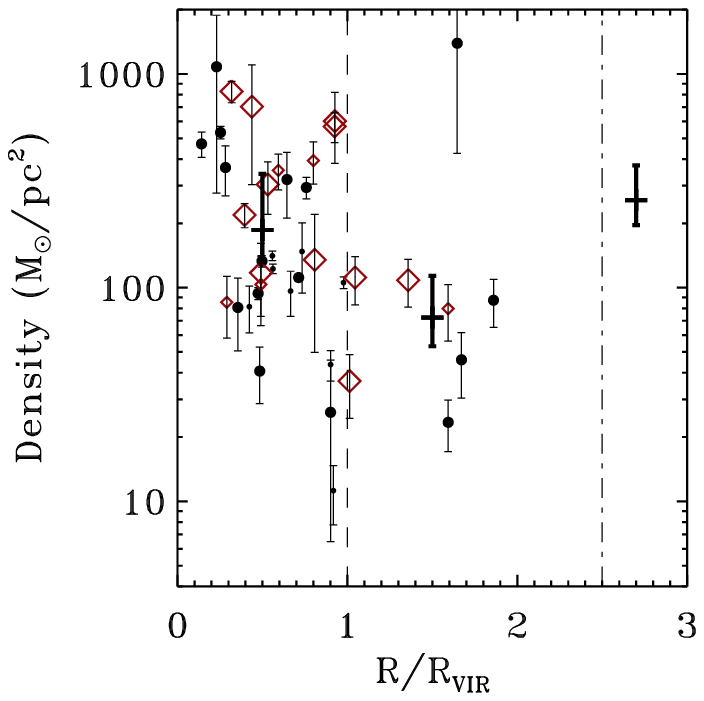}
  \includegraphics*[width=0.95\columnwidth]{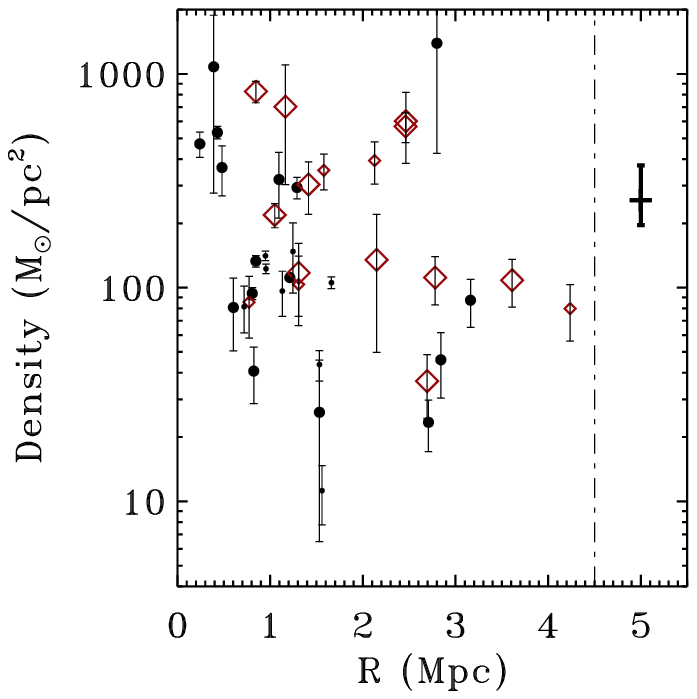}
  \caption{Left: Projected central density of cluster spirals, measured
  within $1.25 r_s$, versus projected $R/R_{VIR}$, calculated
  according to $\Sigma_m(<r)=
  V^2/ (G r)$ Right: Central density versus radius R. 
  In both panels, The thick cross to the right of the dash-dotted line
  indicates the median density for our field sample. In the left
  panel, additional thick crosses indicate the median densities inside
  and outside 1~$R_{VIR}$. Symbols are coded as in Figure~\ref{combinedTF}.}
  \label{densityrvir-fig}
\end{figure*}

\subsection{Tully Fisher Residuals and M/L}

We first turn toward the simplest quantity to examine: the residuals 
from the TF, considered to
represent a change in mass to light ratio (M/L). 
The higher scatter seen in the cluster TF, compared to the field, 
could plausibly be due to changes in star formation rate or dust obscuration 
as galaxies fall into the cluster. This could cause a radial 
gradient in $M/L$, which we would observe as an increased TF scatter.

However, because enhanced scatters are measured in both $V$ and $K_S$ band 
TF relations, we do not expect that the enhancement can be solely 
attributed to an increase or decrease in star formation or dust during cluster
infall. If such a scenario were the main driver of the scatter, we
would expect that the $V$-band scatter would be higher than the
$K_S$-band, yet they are broadly equivalent. We would also
expect to see a correlation between TF residuals and other indicators
of star formation rate or dustiness.  To test this, we plot, in 
Figure~\ref{diffTFcolor}, the V-band TF residuals 
versus $V-K_S$ color for all galaxies in our cluster sample. Here, and
for the rest of this paper, residuals are plotted in the sense that
a positive residual is overluminous for its measured velocity, or,
alternatively, has an anomalously low $V_{MAX}$ given its luminosity.
No obvious correlation is observed in Figure~\ref{diffTFcolor}. 
While several of our objects with lowest
measured $V_{MAX}$ are not plotted here because they do not have
$K_S$ detections, these were also excluded from the TF relation
fitting, and so are not the source of the enhanced scatter.

Even though star formation rate and dust content do not appear to 
correlate with TF residuals, the galaxies of our cluster sample span
a wide range of environments, and so we also examine whether the
increased TF scatter is related to a gradient in $M/L$ across the
cluster, as might be the case if spirals in the outskirts have
formed more recently than those in the cluster cores. 
In Figure~\ref{diffTF}, we plot
the TF residuals as a function of $R/R_{VIR}$, the projected cluster-centric
radius scaled by each cluster's Virial radius: 1.7Mpc for Cl~0024
\citep{tt03} and 2.7 Mpc for MS~0451 \citep{moran06}. As MS~0451 is
much more massive than Cl~0024, a galaxy at 1~Mpc radius in Cl~0024
experiences a very different environment from a galaxy at the same
radius in MS~0451. As we will see below, some key trends emerge 
when we choose to scale by Virial radius, rather than plotting
radius directly on the x-axis. 

Examined by eye, the residuals from the V band TF in 
Figure~\ref{diffTF} (right panel) hint at a radial gradient 
as a function of $R/R_{VIR}$, with galaxies at higher radius seeming to 
be overluminous. However, straight line fits to the $V$ and $K_S$ band residuals find
a small gradient toward higher radius, but with slope no greater than
the error bar on a typical point. A simple gradient in star formation rate
or $M/L$ across the cluster therefore cannot be the only mechanism 
responsible for the increased
scatter in the cluster TF compared to the field. We note, however,
that we cannot rule out the possibility that several different mechanisms are
simultaneously contributing to the TF scatter by acting on spirals in 
different environments within the clusters.

\subsection{Densities and Masses}

Since variations in star formation rate and dust content alone cannot
account for the observed scatter in the cluster TF, 
we are led to consider the idea that the cluster 
spirals are more kinematically disturbed than their field 
counterparts. 
One way to test for disturbed dynamics in a spiral galaxy is to
consider the photometric effective radius, $r_s$, of each
galaxy. We can combine $r_s$ with $V_{MAX}$ to calculate two
fundamental dynamical properties of spiral disks: dynamical mass,
$M(<r)\propto V^2 r$ and central surface mass density, $\Sigma_m(<r) \propto
V^2/r$. Unlike the Fundamental Plane of ellipticals, the Tully-Fisher
relation in the  local universe does not seem to have any dependence
on galaxy size ($r_s$) \citep[e.g.][]{verh01}. 
Therefore, in an undisturbed population of
spirals, we would not expect to uncover any independent environmental
trends in quantities that only depend on $r_s$ and $V_{MAX}$. Any
trends that do exist must be the result of some cluster-related
physical process.

In fact, surface densities allow us to directly probe 
for the action of a key physical mechanism, galaxy harassment 
\citep{moore99}. Harassment is predicted to have a stronger effect
on the least dense galaxies falling into a cluster, to the point of
completely disrupting the most tenuous spirals. Any observed
gradient in the mean density of spirals could then implicate the
action of this physical process.

We choose to study M and $\Sigma_m$ within a radius of $1.25 r_s$, a 
characteristic radius chosen because it is typically 
reached in all of our observed rotation curves, and it is a radius 
at which most of our rotation curves have already leveled off to $V_{MAX}$.

\begin{figure}[t]
  \includegraphics[width=0.95\columnwidth]{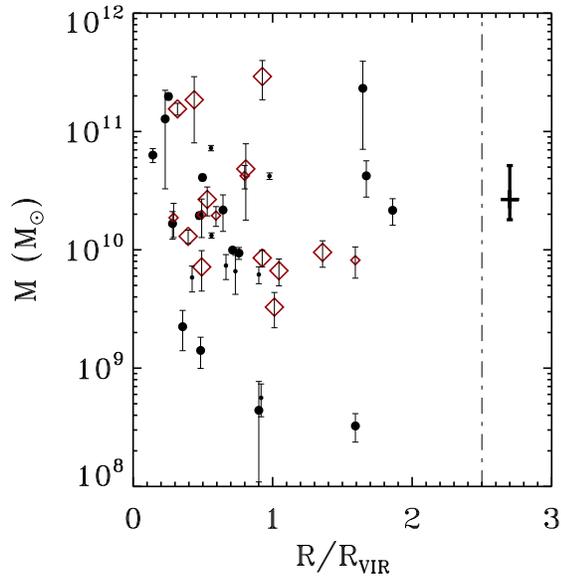}
  \caption{Dynamical mass (within 1.25$r_s$) as a function of
  $R/R_{VIR}$, calculated according to $M(<r)=V^2 r / G$. We find galaxies with a range of dynamical masses, both
  within and outside of the Virial radius. Symbols are coded as in 
  Figure~\ref{combinedTF}. The median for the field sample is
  displayed to the right of the dash-dotted line.}
  \label{massrvir-fig}
\end{figure}

In Figure~\ref{densityrvir-fig}, we plot galaxy density as a function
of projected radius R (right panel) as well as the normalized quantity
$R/R_{VIR}$ (left). In the left panel, one notices a striking break in the
densities of spirals at approximately $1 R_{VIR}$. Near and outside of this
radius, spirals seem to exhibit nearly uniformly low central
densities, which are puzzlingly even lower than those of field
galaxies in the sample. Within the Virial radius, on the other hand, 
a large spread
in densities is seen, and perhaps a radial gradient of decreasing
density outward from the cluster center. In the right hand panel, this trend
appears scrambled, indicating that whatever physical process may cause
this effect, its strength scales as the cluster Viral radius. This
observation rules out several possible mechanisms, and will be discussed
further in the next section.

It is natural to wonder if the observed break in density as a function
of radius is due to simple luminosity segregation: if more massive
spirals are found near the cluster center, perhaps these also have
higher central densities. However, by consulting
Figure~\ref{massrvir-fig}, where
we plot dynamical mass as a function of $R/R_{VIR}$, it becomes
apparent that this is not the case. The low density spirals found in
the clusters' outskirts in fact exhibit a wide spread in total
mass. Because the total number of observed spirals in the cluster
outskirts is small, K-S tests comparing the distributions of mass and
density  for low-R vs. high-R spirals are inconclusive. However, the
cluster sample as a whole does exhibit a larger overall spread in
$\Sigma_m$ than the field sample, and includes a larger fraction of
both low and high density spirals: a K-S test gives a $1.3\%$ chance
that the two samples are drawn from the same parent distribution. This
indicates that the cluster environment may be affecting the internal
mass distributions of spirals at all cluster radii.

\section{DISCUSSION}

What physical mechanisms, then, could be acting on cluster spirals to
reproduce both the overall higher scatter in the TF and the observed
radial trend in density?  The effects seem to persist as far as
$2R_{VIR}$ from the cluster cores, so even though some of the observed galaxies 
at high radius may be part of a ``backsplash population'', it is
very unlikely that nearly all star forming spirals 
in our sample have already been through the cluster center. Therefore,
we do not think it likely that tidal processes are responsible, as
they are only strong near the cluster center. Instead, we can 
consider several proposed 
physical mechanisms that are strong enough at large radius 
to alter the dynamics of a spiral, either directly or indirectly.

In the cluster outskirts, recent mergers are an obvious candidate to
drive both the large TF scatter and the abnormally low densities
\citep[e.g.][and references therein]{tt03} Mergers are less
common in the low density field, and this could explain why we
do not see any low-density spirals in our field sample. Unknown
selection biases, though, could prevent us from including similar
field galaxies in our Tully-Fisher sample.  If the effects of a recent
merger last for at least 1~Gyr \citep{bekki98}, then recent mergers can
affect the TF scatter even in the cluster core \citep{tt03}. 
Therefore, we can not rule out the possibility that increased merging 
in the cluster outskirts serves to drive a high fraction of cluster 
spirals away from the TF relation.

In the inner regions of galaxy clusters, mergers are suppressed due to
the high relative speeds of galaxies, which prevent the creation of a
gravitationally bound pair during close encounters. Instead, an
infalling cluster galaxy is likely to experience repeated close
encounters at high speed due to the high density of galaxies in the
cluster. This process, called galaxy--galaxy harassment, can lead to
dramatic changes in a galaxy.  \citet{moore99} have
shown through simulations that the fate of a harassed galaxy depends
on its original mass and central density. Strongly concentrated Sa/Sb
type galaxies were seen to puff up their disks during infall, and so
harassment may represent one way in which spirals transform into S0s
in clusters. On the other hand, lower density Sc/Sd spirals are more
strongly affected by harassment; \citet{moore99} found that they were
either completely disrupted, or else transformed into an object
resembling a dwarf galaxy.

If harassment is acting to transform the lowest density spirals into
dwarfs, then we would expect to observe a deficit of such low-density
spirals near the cluster cores. High-density spirals, on the other
hand, should be more resistant to harassment, and are likely to
persist to smaller cluster radii. This prediction qualitatively matches the
trend in densities seen in Figure~\ref{densityrvir-fig}, but
the picture is unclear. The puzzling lack of high-density 
galaxies at large cluster radius, and the persistence of low-density
galaxies to $\sim0.5R_{VIR}$ raise questions about
this interpretation. We have already seen in Figure~\ref{massrvir-fig} that
galaxies of a wide range of masses are represented in the cluster
outskirts, so harassment alone may not present a complete explanation
for the observations. Frequent mergers in the cluster outskirts,
however, could very well provide the missing ingredient for keeping
spiral densities low there.

Finally, we consider processes that depend on the hot intracluster
medium (ICM). Generally, even strong interactions with the ICM like
ram pressure stripping are thought to be too weak to explain the 
observed disruptions in the kinematics of spiral disks 
\citep{quilis00}. Rather, such ICM-related processes act largely to
suppress star formation within infalling disks. Since changes in star
formation rate do not appear to be responsible for the increased
scatter in the cluster TF, it is unlikely that an ICM-related process
is involved. 
However, it is possible that shock fronts within the ICM can enhance the ICM's 
ability to affect a spiral disk even at high cluster radius 
\citep[][and references therein]{moran05}. 
Cl~0024 may be undergoing a face-on merger with a large 
group \citep{czoske02}, and so shocks may be important in
this cluster. 
Shocks in the ICM may induce centrally-concentrated 
starbursts within infalling cluster galaxies \citep{moran05}, but it is
not clear that such an interaction would generate emission lines with
enough spatial extent to allow measurement of rotation
curves. Further, since all ICM related processes suppress star
formation over time, our sample of exclusively star-forming 
spirals can only provide and incomplete picture at best of the effects 
of ICM shocks.

One possible concern with our result on the cluster TF relation is
that cluster to cluster variation could be high (as seen for example
by the MORPHS and EDIScS studies, \citep{poggianti99,white05}). 
In fact, previous studies of
Cl~0024 have shown that its galaxy population may be overly active,
possibly due to the ongoing merger with a foreground large group
\citep{czoske02}. Indeed, \citet{moran06} showed that the Fundamental
Plane of elliptical and S0 galaxies in Cl~0024 exhibits a higher
scatter than found in most other intermediate redshift clusters
\citep[e.g.]{kelson00}, though this effect is most significant in the
inner 1~Mpc of the cluster.  It is possible that the increased TF
scatter we see is connected to the similarly enhanced Fundamental
Plane scatter. However, while our cluster TF includes a majority of
points from Cl~0024 (24 galaxies c.f. 16 for MS~0451), it seems clear
by inspection of Figure~\ref{combinedTF} that MS~0451 also contains
spirals that deviate highly from the local TF. As MS~0451 is thought
to be in a more advanced stage of cluster assembly than Cl~0024 (Moran
et al. 2007, in preparation), the universality of our measured TF
scatter remains uncertain until similarly large samples for
several more intermediate redshift clusters become available.

\section{CONCLUSIONS}

In this paper, we have studied the dynamics of cluster and field
spirals at intermediate redshifts, via an analysis of their optical
rotation curves.

We have presented one of the most complete Tully Fisher relations
available for both
cluster and field spirals at these redshifts, and have demonstrated
that the field relation is quite tight even at these high look back
times. In contrast, the cluster TF exhibits a remarkably high
scatter. By comparing the trends of Tully-Fisher residuals vs
radius with those in colors and local mass density we found that the
increased scatter cannot be explained solely in terms of environmental
effects on the star formation rates of infalling galaxies. We
therefore proposed that the increased scatter in the Tully Fisher relation is
due to kinematic disturbances, as expected for example for cluster
harassment.

We also found a trend in galaxy mass density as a function
of cluster centric radius in the sense that spiral galaxies are denser
in the central regions of clusters. This is expected if harassment
plays a significant role, as the densest disks would be most likely to
survive during the infall. However, we found a paucity of high density
spiral galaxies in the cluster outskirts, which cannot be explained by
harassment alone. We suggest that a combination of enhanced merging in
the cluster outskirts with galaxy harassment in the intermediate and
inner cluster regions may be required to explain the observed trend in
galaxy density.

Larger matched samples covering a larger number of galaxy clusters are
needed to determine if the observed trends are universal across
clusters at these redshifts.

\acknowledgements
We thank Taddy Kodama and the PISCES collaboration for kindly making
their Suprime-Cam observations available to us.
We thank Daisuke Nagai, Andrew Benson, and Kevin Bundy for valuable
discussions. GPS thanks Dan Stark, Dave Thompson and Chris Conselice 
for assistance with the observations from Palomar Observatory.
Faint object spectroscopy at Keck Observatory is made possible by
the dedicated effort of the instrument teams and support staff. The 
authors wish to recognize and acknowledge the very significant
cultural role and reverence that the summit of Mauna Kea has always 
had within the indigenous Hawaiian community.  We are most fortunate 
to have the opportunity to conduct observations from this mountain.
GPS acknowledges financial support from a Royal Society 
University Research Fellowship. The analysis pipeline used to reduce the
DEIMOS data was developed at UC Berkeley with support from
NSF grant AST-0071048. We acknowledge support from STScI grants 
HST-GO-08559.01-A and HST-GO-09836.01-A.

\clearpage
\LongTables
\begin{deluxetable}{cccccccccccc}
  \centering
  \tablewidth{0pt}
  \tablecaption{Information and Measurements of Cl~0024, MS~0451
   and field galaxies}
  \tabletypesize{\tiny}

  \tablehead{\colhead{Object} & \colhead{Sample} & \colhead{$z$} & \colhead{RA} & \colhead{DEC} & 
    \colhead{$r_s$} & \colhead{$i$} & \colhead{Sl PA} & 
    \colhead{$M_V$} & \colhead{$M_K$} & \colhead{$V_\textrm{MAX}$} & \colhead{Q}\\
    \colhead{} & \colhead{} & \colhead{} & \colhead{[deg]} & \colhead{[deg]} & 
    \colhead{[kpc]} & \colhead{[deg]} & \colhead{[deg]} &
    \colhead{[mag]} & \colhead{[mag]} & \colhead{[km/s]} & \colhead{}\\
    \colhead{(1)} & \colhead{(2)} & \colhead{(3)} & \colhead{(4)} & \colhead{(5)} & 
    \colhead{(6)} & \colhead{(7)} & \colhead{(8)} &
    \colhead{(9)} & \colhead{(10)} & \colhead{(11)} & \colhead{}}
  \startdata
p0i27c3 &
  Cl0024 &
  0.392 &
  6.639747 &
  17.156639 &
  5.23 &
  54.2 &
  11 & 
  -21.80$\pm$0.05 &
  -22.58$\pm$0.06 &
  219 $\pm$ 14 & 1 \\
p38i14c3 &
  Cl0024 &
  0.400 &
  6.631994 &
  17.154470 &
  4.91 &
  50.5 &
  60 & 
  -21.67$\pm$0.06 &
  -22.32$\pm$0.06 &
  322 $\pm$ 119 & 1 \\
p0i163c4 &
  Cl0024 &
  0.394 &
  6.674678 &
  17.164650 &
  8.70 &
  62.0 &
  3 & 
  -22.89$\pm$0.09 &
  -23.55$\pm$0.05 &
  300 $\pm$ 10 & 1 \\
p0i145c4 &
  Cl0024 &
  0.399 &
  6.676911 &
  17.158810 &
  3.05 &
  28.2 &
  13 & 
  -20.43$\pm$0.07 &
  -20.05$\pm$0.30 &
  147 $\pm$ 19 & 1 \\
N36671 &
  MS0451 &
  0.527 &
  73.521355 &
  -2.994348 &
  6.68 &
  39.6 &
  36 & 
  -22.89$\pm$0.07 &
  -23.47$\pm$0.05 &
  105 $\pm$ 16 & 2 \\
N37826 &
  MS0451 &
  0.530 &
  73.523247 &
  -2.988543 &
  6.18 &
  57.4 &
  10 & 
  -23.12$\pm$0.06 &
  -23.75$\pm$0.05 &
  316 $\pm$ 18 & 1 \\
p38i122c2 &
  Cl0024 &
  0.390 &
  6.618957 &
  17.155710 &
  2.38 &
  36.1 &
  41 & 
  -19.84$\pm$0.06 &
  -19.84$\pm$0.32 &
  61 $\pm$ 11 & 1 \\
N35977 &
  MS0451 &
  0.531 &
  73.506584 &
  -2.993443 &
  3.48 &
  67.4 &
  13 & 
  -20.30$\pm$0.07 &
  -- &
  122 $\pm$ 7 & 1 \\
p37i110c3 &
  Cl0024 &
  0.377 &
  6.671664 &
  17.129601 &
  3.82 &
  80.4 &
  22 & 
  -20.81$\pm$0.07 &
  -- &
  78 $\pm$ 9 & 2 \\
P51837 &
  MS0451 &
  0.526 &
  73.566147 &
  -3.065880 &
  7.33 &
  62.3 &
  55 & 
  -21.77$\pm$0.08 &
  -22.04$\pm$0.05 &
  317 $\pm$ 90 & 1 \\
p37i18c3 &
  Cl0024 &
  0.392 &
  6.676580 &
  17.128189 &
  6.50 &
  69.8 &
  0 & 
  -20.75$\pm$0.05 &
  -- &
  109 $\pm$ 3 & 1 \\
p36i29c3 &
  Cl0024 &
  0.400 &
  6.616654 &
  17.189550 &
  2.66 &
  79.8 &
  29 & 
  -20.62$\pm$0.07 &
  -- &
  45 $\pm$ 6 & 1 \\
N25094 &
  MS0451 &
  0.535 &
  73.499481 &
  -3.053076 &
  6.24 &
  68.7 &
  42 & 
  -21.56$\pm$0.07 &
  -21.61$\pm$0.06 &
  112 $\pm$ 20 & 2 \\
P62230 &
  MS0451 &
  0.530 &
  73.562500 &
  -3.073720 &
  3.53 &
  55.2 &
  44 & 
  -21.21$\pm$0.06 &
  -21.72$\pm$0.05 &
  89 $\pm$ 16 & 1 \\
p36i77c4 &
  Cl0024 &
  0.396 &
  6.640263 &
  17.205460 &
  7.91 &
  77.6 &
  0 & 
  -22.22$\pm$0.06 &
  -22.02$\pm$0.10 &
  143 $\pm$ 4 & 1 \\
P63232 &
  MS0451 &
  0.538 &
  73.494949 &
  -2.982716 &
  4.23 &
  47.4 &
  28 & 
  -20.92$\pm$0.06 &
  -21.35$\pm$0.06 &
  158 $\pm$ 21 & 1 \\
p37i2c3 &
  Cl0024 &
  0.377 &
  6.676347 &
  17.117929 &
  10.23 &
  75.5 &
  0 & 
  -21.31$\pm$0.13 &
  -21.76$\pm$0.20 &
  167 $\pm$ 4 & 2 \\
p36i37c4 &
  Cl0024 &
  0.388 &
  6.630331 &
  17.208891 &
  4.69 &
  69.2 &
  1 & 
  -21.07$\pm$0.06 &
  -20.65$\pm$0.23 &
  105 $\pm$ 2 & 2 \\
P54663 &
  MS0451 &
  0.541 &
  73.597534 &
  -3.063867 &
  3.35 &
  56.1 &
  21 & 
  -21.65$\pm$0.06 &
  -22.35$\pm$0.05 &
  152 $\pm$ 14 & 2 \\
p19i2c4 &
  Cl0024 &
  0.396 &
  6.591903 &
  17.168131 &
  3.71 &
  31.0 &
  29 & 
  -21.47$\pm$0.06 &
  -22.62$\pm$0.05 &
  152 $\pm$ 26 & 1 \\
p19i1c4 &
  Cl0024 &
  0.397 &
  6.595224 &
  17.186880 &
  3.94 &
  32.8 &
  8 & 
  -20.69$\pm$0.07 &
  -20.80$\pm$0.20 &
  86 $\pm$ 10 & 2 \\
p13i19c4 &
  Cl0024 &
  0.381 &
  6.617717 &
  17.218800 &
  4.26 &
  78.8 &
  0 & 
  -21.34$\pm$0.05 &
  -- &
  96 $\pm$ 1 & 1 \\
p23i151c2 &
  Cl0024 &
  0.393 &
  6.679020 &
  17.103390 &
  3.02 &
  23.8 &
  21 & 
  -20.32$\pm$0.05 &
  -22.09$\pm$0.06 &
  93 $\pm$ 16 & 2 \\
p18i234c4 &
  Cl0024 &
  0.396 &
  6.721228 &
  17.167561 &
  2.55 &
  62.9 &
  16 & 
  -20.14$\pm$0.08 &
  -20.25$\pm$0.29 &
  121 $\pm$ 6 & 1 \\
N27099 &
  MS0451 &
  0.550 &
  73.635056 &
  -3.040075 &
  4.67 &
  77.9 &
  29 & 
  -21.73$\pm$0.06 &
  -21.35$\pm$0.07 &
  189 $\pm$ 21 & 2 \\
N49410 &
  MS0451 &
  0.528 &
  73.533180 &
  -2.924410 &
  8.54 &
  35.4 &
  53 & 
  -21.70$\pm$0.05 &
  -21.93$\pm$0.05 &
  150 $\pm$ 47 & 1 \\
p24i55c2 &
  Cl0024 &
  0.388 &
  6.639475 &
  17.082970 &
  1.85 &
  71.6 &
  27 & 
  -19.29$\pm$0.09 &
  -- &
  30 $\pm$ 11 & 1 \\
p25i136c3 &
  Cl0024 &
  0.395 &
  6.584536 &
  17.114901 &
  5.37 &
  81.9 &
  24 & 
  -21.16$\pm$0.08 &
  -- &
  67 $\pm$ 5 & 2 \\
p24i1c2 &
  Cl0024 &
  0.392 &
  6.646124 &
  17.081320 &
  3.19 &
  60.4 &
  13 & 
  -19.93$\pm$0.10 &
  -20.85$\pm$0.24 &
  26 $\pm$ 4 & 2 \\
N46608 &
  MS0451 &
  0.540 &
  73.473412 &
  -2.938749 &
  1.75* &
  61.1 &
  21 & 
  -21.79$\pm$0.05 &
  -22.67$\pm$0.05 &
  139 $\pm$ 13 & 1 \\
N48527 &
  MS0451 &
  0.544 &
  73.488052 &
  -2.927898 &
  9.94 &
  50.3 &
  40 & 
  -22.05$\pm$0.05 &
  -22.60$\pm$0.05 &
  341 $\pm$ 62 & 1 \\
p13i15c2 &
  Cl0024 &
  0.397 &
  6.598421 &
  17.232731 &
  9.00 &
  75.9 &
  0 & 
  -21.29$\pm$0.05 &
  -- &
  136 $\pm$ 4 & 2 \\
N34128 &
  MS0451 &
  0.534 &
  73.427948 &
  -3.003162 &
  4.28 &
  75.5 &
  3 & 
  -20.84$\pm$0.06 &
  -- &
  55 $\pm$ 9 & 1 \\
N48766 &
  MS0451 &
  0.551 &
  73.469215 &
  -2.925409 &
  3.49 &
  33.9 &
  26 & 
  -21.30$\pm$0.05 &
  -22.14$\pm$0.05 &
  87 $\pm$ 11 & 1 \\
N47732 &
  MS0451 &
  0.534 &
  73.413109 &
  -2.930732 &
  4.23 &
  46.3 &
  24 & 
  -20.26$\pm$0.06 &
  -20.17$\pm$0.15 &
  94 $\pm$ 11 & 1 \\
N64729 &
  MS0451 &
  0.540 &
  73.520599 &
  -2.835176 &
  4.57 &
  44.2 &
  38 & 
  -21.37$\pm$0.06 &
  -21.80$\pm$0.05 &
  84 $\pm$ 12 & 2 \\
p7i72c2 &
  Cl0024 &
  0.373 &
  6.597922 &
  17.299730 &
  1.68 &
  46.0 &
  18 & 
  -19.04$\pm$0.07 &
  -- &
  27 $\pm$ 3 & 1 \\
p30i4c3 &
  Cl0024 &
  0.393 &
  6.552849 &
  17.050751 &
  5.82 &
  35.8 &
  54 & 
  -21.46$\pm$0.13 &
  -22.48$\pm$0.10 &
  398 $\pm$ 138 & 1 \\
p10i41c4 &
  Cl0024 &
  0.396 &
  6.801073 &
  17.199739 &
  13.68 &
  41.4 &
  30 & 
  -22.30$\pm$0.10 &
  -23.04$\pm$0.05 &
  110 $\pm$ 18 & 1 \\
p33i308c4 &
  Cl0024 &
  0.382 &
  6.603914 &
  17.000919 &
  7.10 &
  46.8 &
  26 & 
  -20.21$\pm$0.05 &
  -20.65$\pm$0.25 &
  110 $\pm$ 13 & 1 \\
p5i171c4 &
  field &
  0.313 &
  6.742421 &
  17.238001 &
  3.43 &
  74.8 &
  0 & 
  -20.77$\pm$0.11 &
  -21.38$\pm$0.07 &
  99 $\pm$ 3 & 1 \\
N33315 &
  field &
  0.314 &
  73.373642 &
  -3.008564 &
  4.16 &
  55.8 &
  47 & 
  -21.67$\pm$0.13 &
  -- &
  159 $\pm$ 29 & 1 \\
N22367 &
  field &
  0.325 &
  73.610954 &
  -3.070553 &
  5.06 &
  74.7 &
  45 & 
  -21.95$\pm$0.05 &
  -22.29$\pm$0.05 &
  185 $\pm$ 34 & 1 \\
N47194 &
  field &
  0.326 &
  73.525711 &
  -2.939083 &
  7.08 &
  33.0 &
  44 & 
  -21.81$\pm$0.06 &
  -22.27$\pm$0.05 &
  172 $\pm$ 43 & 1 \\
N58982 &
  field &
  0.326 &
  73.631409 &
  -2.867468 &
  4.26 &
  44.3 &
  14 & 
  -20.04$\pm$0.06 &
  -19.57$\pm$0.17 &
  115 $\pm$ 10 & 1 \\
N11649 &
  field &
  0.329 &
  73.513878 &
  -3.123671 &
  2.40 &
  57.4 &
  42 & 
  -20.53$\pm$0.07 &
  -21.36$\pm$0.05 &
  80 $\pm$ 14 & 1 \\
N30997 &
  field &
  0.333 &
  73.444008 &
  -3.020530 &
  2.59 &
  62.4 &
  20 & 
  -20.55$\pm$0.06 &
  -21.05$\pm$0.05 &
  104 $\pm$ 8 & 1 \\
N11932 &
  field &
  0.362 &
  73.463303 &
  -3.124372 &
  3.95 &
  41.0 &
  67 & 
  -20.96$\pm$0.06 &
  -21.27$\pm$0.05 &
  235 $\pm$ 133 & 1 \\
N48819 &
  field &
  0.363 &
  73.460930 &
  -2.927250 &
  1.93 &
  28.1 &
  26 & 
  -20.41$\pm$0.05 &
  -21.16$\pm$0.05 &
  83 $\pm$ 10 & 2 \\
N55288 &
  field &
  0.364 &
  73.419220 &
  -2.892889 &
  5.20 &
  62.6 &
  50 & 
  -20.56$\pm$0.06 &
  -20.23$\pm$0.08 &
  113 $\pm$ 17 & 2 \\
N37713 &
  field &
  0.367 &
  73.467995 &
  -2.987958 &
  4.95 &
  27.2 &
  0 & 
  -21.06$\pm$0.05 &
  -21.91$\pm$0.05 &
  204 $\pm$ 20 & 1 \\
N21079 &
  field &
  0.371 &
  73.350967 &
  -3.074297 &
  2.86 &
  42.9 &
  33 & 
  -20.45$\pm$0.05 &
  -21.42$\pm$0.05 &
  118 $\pm$ 18 & 1 \\
N31950 &
  field &
  0.371 &
  73.444237 &
  -3.017617 &
  3.69 &
  55.2 &
  41 & 
  -21.43$\pm$0.05 &
  -22.03$\pm$0.05 &
  182 $\pm$ 32 & 1 \\
N35309 &
  field &
  0.390 &
  73.625328 &
  -2.996757 &
  3.32 &
  66.3 &
  21 & 
  -21.18$\pm$0.05 &
  -21.51$\pm$0.05 &
  131 $\pm$ 11 & 1 \\
N57426 &
  field &
  0.391 &
  73.473892 &
  -2.878548 &
  3.73 &
  60.3 &
  38 & 
  -21.66$\pm$0.05 &
  -22.98$\pm$0.05 &
  274 $\pm$ 43 & 1 \\
N18456 &
  field &
  0.401 &
  73.598076 &
  -3.087556 &
  4.13 &
  61.0 &
  14 & 
  -21.55$\pm$0.05 &
  -22.13$\pm$0.05 &
  182 $\pm$ 11 & 1 \\
N35519 &
  field &
  0.413 &
  73.426918 &
  -2.997052 &
  3.83 &
  40.8 &
  43 & 
  -20.43$\pm$0.05 &
  -20.76$\pm$0.06 &
  135 $\pm$ 27 & 2 \\
N19992 &
  field &
  0.419 &
  73.515671 &
  -3.081446 &
  6.52 &
  51.5 &
  4 & 
  -21.36$\pm$0.20 &
  -- &
  121 $\pm$ 11 & 1 \\
N18483 &
  field &
  0.425 &
  73.465881 &
  -3.088012 &
  4.56 &
  71.7 &
  3 & 
  -22.18$\pm$0.05 &
  -22.83$\pm$0.05 &
  177 $\pm$ 6 & 1 \\
p17i129c4 &
  field &
  0.443 &
  6.766013 &
  17.151690 &
  10.24 &
  84.6 &
  2 & 
  -22.69$\pm$0.07 &
  -21.99$\pm$0.20 &
  215 $\pm$ 3 & 2 \\
N36846 &
  field &
  0.447 &
  73.670662 &
  -2.991703 &
  2.73 &
  45.3 &
  0 & 
  -21.60$\pm$0.05 &
  -22.75$\pm$0.05 &
  180 $\pm$ 11 & 1 \\
N40959 &
  field &
  0.447 &
  73.624619 &
  -2.969724 &
  3.40 &
  34.8 &
  13 & 
  -22.12$\pm$0.05 &
  -22.88$\pm$0.05 &
  187 $\pm$ 19 & 1 \\
N41465 &
  field &
  0.447 &
  73.629173 &
  -2.963893 &
  4.67 &
  74.7 &
  21 & 
  -21.33$\pm$0.05 &
  -20.84$\pm$0.08 &
  142 $\pm$ 11 & 1 \\
N56286 &
  field &
  0.463 &
  73.505577 &
  -2.884414 &
  4.59 &
  53.6 &
  3 & 
  -21.53$\pm$0.05 &
  -21.70$\pm$0.05 &
  209 $\pm$ 10 & 1 \\
p11i11c4 &
  field &
  0.476 &
  6.724503 &
  17.193130 &
  2.77 &
  27.3 &
  18 & 
  -20.51$\pm$0.09 &
  -20.69$\pm$0.27 &
  74 $\pm$ 13 & 1 \\
N25636 &
  field &
  0.491 &
  73.394081 &
  -3.050711 &
  6.04 &
  28.5 &
  15 & 
  -21.67$\pm$0.06 &
  -22.66$\pm$0.05 &
  174 $\pm$ 22 & 2 \\
p8i163c3 &
  field &
  0.492 &
  6.569071 &
  17.296320 &
  3.14 &
  56.7 &
  7 & 
  -20.97$\pm$0.10 &
  -21.55$\pm$0.19 &
  128 $\pm$ 5 & 1 \\
N59564 &
  field &
  0.494 &
  73.547920 &
  -2.870288 &
  4.66 &
  71.4 &
  22 & 
  -22.00$\pm$0.05 &
  -22.16$\pm$0.05 &
  172 $\pm$ 15 & 1 \\
p1i97c4 &
  field &
  0.494 &
  6.770693 &
  17.324760 &
  3.00* &
  38.8 &
  16 & 
  -21.59$\pm$0.06 &
  -22.20$\pm$0.12 &
  59 $\pm$ 8 & 2 \\
N11338 &
  field &
  0.505 &
  73.540504 &
  -3.126556 &
  5.12 &
  36.2 &
  33 & 
  -21.76$\pm$0.06 &
  -22.80$\pm$0.05 &
  258 $\pm$ 45 & 1 \\
N25124 &
  field &
  0.506 &
  73.698349 &
  -3.052146 &
  7.84 &
  39.4 &
  44 & 
  -21.90$\pm$0.06 &
  -22.83$\pm$0.05 &
  194 $\pm$ 45 & 1 \\
p23i180c2 &
  field &
  0.536 &
  6.677425 &
  17.101549 &
  5.56 &
  56.0 &
  14 & 
  -21.26$\pm$0.07 &
  -22.08$\pm$0.14 &
  146 $\pm$ 10 & 1 \\
p21i99c3 &
  field &
  0.537 &
  6.475233 &
  17.202730 &
  2.68 &
  36.1 &
  3 & 
  -20.93$\pm$0.05 &
  -20.97$\pm$0.28 &
  157 $\pm$ 14 & 1 \\
N37427 &
  field &
  0.579 &
  73.516296 &
  -2.988545 &
  5.59 &
  31.0 &
  43 & 
  -22.45$\pm$0.06 &
  -23.48$\pm$0.05 &
  215 $\pm$ 53 & 1 \\
p34i20c3 &
  field &
  0.595 &
  6.533985 &
  16.984360 &
  4.26 &
  53.3 &
  2 & 
  -21.15$\pm$0.05 &
  -21.06$\pm$0.32 &
  109 $\pm$ 7 & 1 \\
N41286 &
  field &
  0.600 &
  73.466621 &
  -2.966479 &
  2.53 &
  43.0 &
  22 & 
  -21.33$\pm$0.05 &
  -20.67$\pm$0.11 &
  107 $\pm$ 12 & 1 \\
p34i94c2 &
  field &
  0.614 &
  6.553948 &
  16.984819 &
  11.47 &
  38.2 &
  30 & 
  -21.23$\pm$0.07 &
  -21.58$\pm$0.33 &
  251 $\pm$ 37 & 2 \\

  \enddata
  \tablecomments{Cluster galaxies are arranged in order of increasing
    $R/R_{VIR}$, where $R_{VIR}=1.70$~Mpc for Cl~0024 and $R_{VIR}=2.66$ for
    MS~0451. The cluster centers are (6.6500,17.1433) and (73.5454,-3.0186) J2000, for Cl~0024 and MS~0451,
    respectively. Field galaxies are arranged in ascending redshift order.
    Col. (1): Object name. Col (2): Indicates which
    subsample that each galaxy belongs to: Cl~0024, MS~0451, or field.
    Col (3): Redshift of each galaxy. Cols (4) and (5): J2000
    coordinates for each galaxy. Cols (6) and (7): Scale lengths $r_s$
    and inclinations $i$ measured via GALFIT. $i=90^\circ$ indicates
    edge-on. * Denotes objects where $R_S$ was set manually to improve
    the fit. Col (8): Indicates the misalignment
    between the PA of the spectroscopic slit and the major axis of the
    galaxy. Typical formal errors on $r_s$, $i$, and Sl
    PA are 10$\%$, $0.5^{\circ}$, and $1^{\circ}$, respectively; 
    systematic uncertainties in $i$
    and Sl PA are typically $\sim10^{\circ}$.
   Cols (9) and (10): Absolute magnitudes in rest-frame V and
    K bands, corrected for inclination-dependent internal
    extinction. Col (11): Measured $V_{MAX}$. In cases where more than
    one emission line was measured, this is a weighted average. Col
    (12): Rotation curve quality. $Q=1$ curves display turnovers on
    both ends; Q=2 curves display only one turnover, or an uncertain
    fit. }
\end{deluxetable}

\end{document}